\documentclass[onecolumn,sort&compress,numbers]{els-mrw} 

\usepackage{amsmath,amssymb,amsfonts,amsthm,makeidx,graphicx}
\usepackage{txfonts}
\usepackage{helvet}
\usepackage{hyperref}
\usepackage{cancel}

\begin{document}

\chapter{Chiral spin symmetry}\label{chap1}

\author[1]{L. Ya. Glozman}

\address[1]{\orgname{Institute of Physics}, \orgdiv{University of Graz}, \orgaddress{Universitaetsplatz 5, A-8010, Graz, Austria}}

%\articletag{Chapter Article tagline: update of previous edition,, reprint..}

\maketitle

%\begin{glossary}[Glossary]
%\term{Europe} the model is a coherent view of capital markets data that allows %users to interact with the content in a consistent manner.

%\term{Primates} regardless of the source. Essentially, of sources. Properly %deployed.

%\end{glossary}

%\begin{glossary}[Nomenclature]
%\begin{tabular}{@{}lp{34pc}@{}}
%AF &Assessment Factor\\
%ECHA &European Chemical Agency\\
%EPM &Equilibrium Partitioning Method Equilibrium Partitioning Method Equilibrium %Partitioning Method Equilibrium\hfill\break Partitioning Method\\
%ERA &Ecological Risk Assessment\\
%HC &Hazardous Concentration\\
%\end{tabular}
%\end{glossary}

\begin{abstract}[Abstract]
We review the recently discovered new symmetry in QCD, the chiral spin symmetry, which is
a symmetry of the color charge and of the confining electric part
of the QCD Lagrangian. Observation of this symmetry on the lattice
in the vacuum upon truncation of the near-zero modes of the Dirac
operator implies that the hadron mass generation in the light quark
sector is not a consequence of the spontaneous breaking of chiral symmetry
and that  confinement and chiral symmetry breaking are not directly related
in QCD. Observation of this symmetry at high temperatures above the
chiral symmetry restoration crossover suggests that QCD is still
in the confining regime and the degrees of freedom are chirally symmetric
quarks bound into the color-singlet systems by the confining electric field.
This regime of QCD was called a stringy fluid.
At a temperature $T_d$ that is essentially above $T_{ch}$ the chiral spin symmetry smoothly disappears suggesting that the confining electric field gets
screened and one observes a very smooth crossover to the quark-gluon plasma.
The three-regimes picture of the QCD phase diagram at small chemical
potentials has been further substantiated by the analysis of the $N_c$
scaling of main thermodynamic observables: the energy density, the pressure
and the entropy density. In the hadron gas regime they scale as $N_c^0$,
in the stringy fluid as $N_c^1$ and in the quark-gluon plasma as $N_c^2$.
This scaling can be observed on the lattice upon
simulations of the equation of state at $N_c>3$. We have analyzed the
fluctuations of conserved charges that  by construction scale as $N_c^1$ 
above $T_{ch}$ thus indicating a transition from the hadron gas to the
stringy fluid. When $N_c$ gets sufficiently large the smooth crossovers
become first order phase transitions and the three-regimes picture 
transforms into the three-phases phase diagram. Finally we discuss
a manifestly confining and chirally symmetric model in 3+1 dimensions, which is similar to the 't Hooft model in 1+1 dimensions. 
This model demonstrates the chiral symmetry restoration in the confining regime and a delocalization of the color-singlet quark-antiquark systems
that become very large at $T > T_{ch}$. This happens because of Pauli 
blocking of the small momenta quark levels by the thermal quark excitations.
The huge swelling of mesons above $T_{ch}$ implies that the stringy fluid matter is a very dense highly collective system of the overlapping very large color-singlet quark-antiquark "mesons" with a very small mean free path.
\end{abstract}

\begin{keywords}
 	chiral spin symmetry \sep QCD phase diagram 
\end{keywords}

\bigskip
\bigskip
\noindent
{\bf \large Key Points/Objectives:}
\medskip
\begin{itemize}%
\item{Chiral spin symmetry and its extensions in QCD Lagrangian and Hamiltonian.}
\item{Observation of the chiral spin symmetry and its extensions in hadrons
at $T=0$ and its implications.}
\item{Observation of the chiral spin symmetry and its extensions at high
temperatures above the chiral restoration crossover and its implications.}
\item{Three regimes/phases of QCD at small chemical potentials, their
symmetries and $N_c$ scaling.}
\item{The origin of the $N_c^1$ scaling in the stringy fluid regime/phase.}
\item{Fluctuations of conserved charges as evidence for the transition from the hadron gas to the stringy fluid.}
\item{Chirally symmetric and confining model. Restoration of chiral symmetry in the confining regime. Properties of hadrons above $T_{ch}$.}
\end{itemize}
%---------------------------------------------------------------

\section{Introduction}\label{sec:intro}

This review is devoted to the recently discovered new symmetry
in QCD, the chiral spin symmetry $SU(2)_{CS}$ and its flavor extensions
 $SU(2N_F)$ and $SU(2N_F) \times SU(2N_F)$, which are symmetries of the electric confining part of QCD with light quarks. We discuss the history, 
  the algebraic structure of these symmetries as well as their physics content. Then we
 address implications of these symmetries for the origin of hadron
 masses, for the absence of the direct interrelation between the
 confinement and  spontaneous  breaking 
 of chiral symmetry in QCD. The second part of the review is devoted to the implications of these symmetries for the QCD phase
 diagram at small baryon chemical potential. In particular we demonstrate
 that the QCD phase diagram is more complicated than was believed in previous
 years and that between the hadron gas regime/phase and the quark-gluon plasma regime/phase
 there is an intermediate regime/phase, named stringy fluid, where chiral
 symmetry is restored but the system is still with confinement
 with the color-singlet objects being the propagating in time degrees of freedom. We address
 the $N_c$ scaling of all three regimes/phases and finally present the confining
 and chirally symmetric model that  sheds the light on the microscopic structure of the intermediate phase.

\section{Unexpected degeneracy}\label{sec:degeneracy}

The history of the chiral spin symmetry begins with the
observation on the lattice of an unexpected degeneracy of the isovector mesons
upon the artificial restoration of chiral symmetry. Author's student Misha Denissenya,
the long term collaborator Christian Lang and the author were interested in the
origin of hadron masses in the light quark sector. Quite general opinion
in the community was that the hadron mass is originated from the spontaneous
breaking of chiral symmetry, which was motivated by a number of models, like the $sigma$-model,
the MIT bag model, the constituent quark model, the Skyrme model  and also by the QCD sum rules.
We asked the following question: what will happen with the hadron mass if we
remove effects related with the spontaneous breaking of chiral symmetry?

The chiral $SU(2)_R \times SU(2)_L$ transformation in a world where the chiral
symmetry  is not broken, connects hadrons that belong to the one and the same
multiplet of the $SU(2)_R \times SU(2)_L$ group. Examples of 
how such transformations are performed, can be found in a review \cite{G1}. An example of the
chiral multiplet are e.g. the pion and the sigma-meson, which transform
 into one another  by the chiral transformation. Or the $\rho$ and $a_1$ mesons,
 which also transform into one another via the $SU(2)_R \times SU(2)_L$.
 In the chirally symmetric world hadrons that belong to a given chiral 
 multiplet must be degenerate. Similarly, the $U(1)_A$ symmetry, if unbroken,
 connects hadrons that transform into one another under the $U(1)_A$
 transformation. An example of the $U(1)_A$ multiplet is e.g. the $\pi - a_0$
 pair. A complete list of the $SU(2)_R \times SU(2)_L$ and $U(1)_A$ multiplets
 of the $J=1$ mesons \cite{CJ} is presented
in the left panel of Fig. 1. Consequently, in the $SU(2)_R \times SU(2)_L$ symmetric world one should expect degenerate pairs of mesons that are connected by the red arrows
on the l.h.s. of Fig. 1. In the $U(1)_A$ symmetric world mesons that are connected by the blue arrows must be degenerate. If both symmetries are 
restored then the degenerate multiplets should consist of mesons that are connected by both the red and blue arrows.

The quark condensate of the vacuum is connected to the density of the
near-zero modes
of the Dirac operator in Euclidean space via the Banks-Casher relation \cite{BC}

\begin{equation}
<\bar q q> = -\pi \rho(0).
\label{bc}
\end{equation}
\noindent
All eigenmodes of the Dirac operator can be obtained from the solution
of the eigenvalue problem 

\begin{equation}
i \gamma_\mu D_\mu  \psi_n(x) = \lambda_n \psi_n(x).
\label{ep}
\end{equation}
\noindent
Consequently, we could remove effects of spontaneous breaking of chiral
symmetry by subtracting the lowest modes of the Dirac operator from the
full quark propagators:
\noindent
\begin{equation}
 S =S_{Full}-
  \sum_{i=1}^{k}\,\frac{1}{\lambda_i}\,|\lambda_i\rangle \langle \lambda_i|. 
\label{tr}
\end{equation}
\noindent
Applying then the standard variational technique, which is used on the lattice
to extract  hadron masses , we could study the evolution
of the hadron masses as a function of the number  $k$ of removed lowest modes
of the Dirac operator. When chiral symmetry is restored and if at the
same time the hadrons survive the truncation procedure,  one a-priori expects
a degeneracy of hadrons which are connected by the chiral
$SU(2)_R \times SU(2)_L$ transformations and, perhaps, a degeneracy
of hadrons that are connected by the $U(1)_A$ transformation (if the
effect of the explicit  $U(1)_A$ breaking is located in the lowest Dirac eigenmodes).
\begin{figure}
\centering
\includegraphics[angle=0,width=0.48\linewidth]{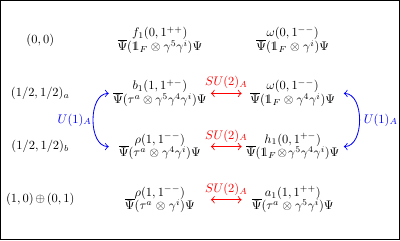}
\includegraphics[angle=0,width=0.48\linewidth]{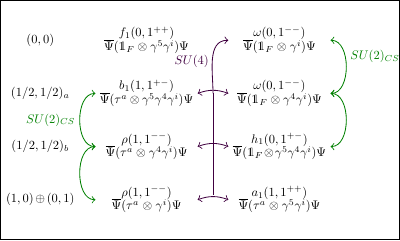}
\label{multiplets1}
\caption{Transformations between $J=1$ operators, $i=1,2,3$. 
Left panel: The left
column indicates the $SU(2)_R \times SU(2)_L$ representations for every operator,
with $(I_R,I_L)$ being the isospins of the right- and left-handed quarks (with
the total isospin $I$ restricted to be $ |I_R - I_L| \leq I \leq I_R + I_L$).
$I, J^{PC}$ together with the chiral representation form a complete
set of quantum numbers of the operator.
Red and blue arrows connect operators which transform into each other under
$SU(2)_R \times SU(2)_L$ and $U(1)_A$, respectively. $\gamma^4 = \gamma^0$.
Right panel: The $J=1$ $SU(2)_{CS}$ triplets (green arrows) and the $SU(4)$
15-plet (purple). The $f_1$ operator is a singlet of $SU(4)$. 
Other notations are the same as in the left panel. From
ref. \cite{GP}.}
\end{figure}
If hadrons survive the artificial chiral
symmetry restoration, then according to  the left panel of Fig. 1
one should expect  a degeneracy of the $\rho$-meson from
the $(1,0)+(0,1)$ representation with the $a_1$-meson,  a degeneracy of the
$\rho$-meson from
the $(1/2,1/2)_b$ representation with the $h_1$-meson, etc.
If at the same time the $U(1)_A$ symmetry is restored (which is broken
by the quark condensate and by the $U(1)_A$ anomaly), one should observe
 a degeneracy of the $b_1$ meson with the  $\rho$-meson from
the $(1/2,1/2)_b$ representation, etc. However some isovector operators 
are not connected
by both transformations and their masses should be expected to be different
after chiral restoration, e.g. masses of $b_1$ and $a_1$ mesons.
\begin{figure}
\centering
\includegraphics[angle=0,width=0.35\linewidth]{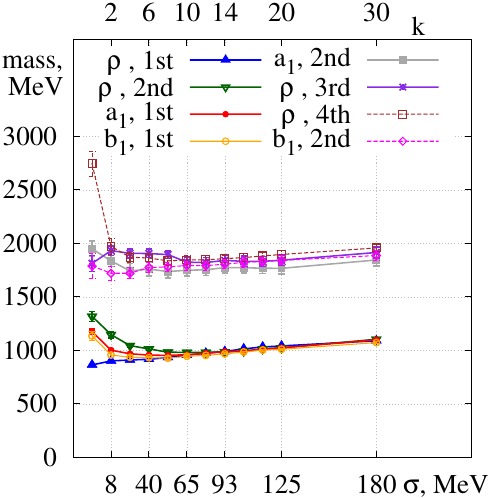}
\label{iso}
\caption{$J=1$ isovector meson masses as a function of the truncation number $k$,
where $k$ represents the amount of removed lowest modes of the Dirac operator. $\sigma$ shows the energy gap in the Dirac spectrum. From ref. \cite{D1}.}
\end{figure}

In Fig. 2 we show the results \cite{D1} of the mass evolution of the isovector $J=1$ mesons. We observed that the mesons do survive the
artificial chiral restoration: the quality of the exponential decay of the
correlators was even essentially better than in the untruncated case. Since the
hadrons survive and at the same time their mass is large in the chirally symmetric world, one concludes that the view that the hadron mass comes
from the spontaneous breaking of chiral symmetry  was erroneous. All
degeneracies required by the $SU(2)_R \times SU(2)_L$ and $U(1)_A$ restorations
are satisfied. However, one observes a larger degeneracy: all isovector mesons
get degenerate. This was completely unexpected:
one observes some symmetry that is larger than the $SU(2)_R \times SU(2)_L \times U(1)_A$ chiral symmetry of QCD!

\section{Chiral spin symmetry in QED and QCD}\label{sec:qedandqcd}
The symmetry group, $SU(4)$, that is responsible for the observed degeneracy, and its physics origin were obtained in Ref. \cite{G2}. The multiplet
structure of the $SU(4)$ group is depicted on the r.h.s. of Fig. 1 \cite{GP}.

Consider Maxwell equations
\begin{eqnarray}
div \vec E &=& 4\pi \rho \nonumber\\
rot \vec B - \frac{1}{c}\frac{\partial \vec E}{\partial t} &=& \frac{4\pi}{c}\vec j\nonumber\\
rot \vec E + \frac{1}{c}\frac{\partial \vec B}{\partial t} &=& 0 \nonumber \\
div \vec B &=& 0 \nonumber
\end{eqnarray} 
\noindent
We define $\vec E$ and $\vec B$ in a given Lorentz frame through their
interaction with a charge and a current:
 
\begin{equation}
\vec F = q \vec E + q \frac{\vec v}{c} \times  \vec B.
\end{equation}
\noindent
Assume charges to be particles with $s = 1/2$. 
They are characterized by helicities (chiralities for massless particles)
and the Dirac bispinor in the chiral (Weyl) representation is
\begin{equation}
\left(\begin{array}{c}
R\\
L
\end{array}\right)\;.
\end{equation}
Consider a $SU(2)_{CS}$ chiral spin transformation that mixes the right- and left-handed Weyl spinors $R$ and $L$ \cite{G2}:
\begin{equation}
\left(\begin{array}{c}
R\\
L
\end{array}\right)\; \rightarrow
\left(\begin{array}{c}
R'\\
L'
\end{array}\right)=
\exp \left(i  \frac{\varepsilon^n \sigma^n}{2}\right) \left(\begin{array}{c}
R\\
L
\end{array}\right)\;. 
\end{equation}\medskip
\noindent
Obviously the Dirac Lagrangian prohibits such transformation.
What happens with the  charge density $\rho$ and the current 
density $ \vec j = \rho \vec v$ upon the chiral spin transformation?

\begin{equation}
R'^\dagger R' + L'^\dagger L' = R^\dagger R + L^\dagger L,
\end{equation}
i.e.
\begin{equation}
\rho' = \rho.
\end{equation}
The charge density is invariant under the chiral spin transformation.
However the  current density $\vec j$ changes because
$\vec v$  changes.
Hence the interaction of a charge with the electric field
is invariant under $SU(2)_{CS}$,
while the interaction of a current with the magnetic field
is not.
We can distinguish the electric and magnetic fields in a given Lorentz
frame 
by the chiral spin symmetry!
The electric part of the EM theory is more symmetric
than the magnetic part. 
Notice that it is a gauge-invariant statement since the fields $\vec E$
and $\vec B$ do not depend on a choice of a particular gauge.

The QED Lagrangian contains both the interaction of the charge with the
electric field and of the spatial current with the magnetic one
\begin{equation}
 {\cal L} ={\cal L}(\vec E, \vec B)
+ \rho \phi - \vec j \cdot \vec A +  ~~ Dirac ~~Lagrangian.
\end{equation}
The Dirac Lagrangian is not invariant under the $SU(2)_{CS}$.
It is the electric part of the Lagrangian,  $\rho\phi$, that is invariant under
$SU(2)_{CS}$. Of course, we can fix a gauge which  eliminates
the scalar potential, $\phi=0$,  which is  the temporal or Weyl gauge.
However this gauge fixing does not eliminate the interaction
between the electric charge and electric field, which is simply 
shifted to the other part of the Lagrangian \cite{Jackson}.

In terms of the Dirac spinors $\psi$ the same $SU(2)_{CS}$
 transformation can be written via $\gamma$-matrices \cite{GP}
\begin{equation}
\label{V-defsp}
  \psi \rightarrow  \psi^\prime = \exp \left(i  \frac{\varepsilon^n \Sigma^n}{2}\right) \psi\; ,
\end{equation}

\noindent
where the generators $\Sigma^n$ of the four-dimensional reducible
representation are

\begin{equation}
 \Sigma^n = \{\gamma_0,-i \gamma_5\gamma_0,\gamma_5\}, ~~~[\Sigma^a,\Sigma^b]=2i\epsilon^{abc}\Sigma^c.
\label{SIGCS}
\end{equation}

\noindent
The $U(1)_A$ group is a subgroup of $SU(2)_{CS}$.

\medskip
Now we turn to QCD, where things are very similar.
The QCD equation of motion prescribes the interaction of
the Lorentz-invariant color charge of quarks, 
\begin{equation}
Q^a = g\int d^3x ~ 
\psi^\dagger(x) T^a \psi(x),
\end{equation}
where $T^a,~~a=1,...,8$ are the $SU(3)$ color generators,
with the chromoelectric
field $\vec E^a, ~~~~a=1,...,8$,
to be of the same form as in electrodynamics:
\begin{equation}
 \vec F = Q^a \vec E^a.
\end{equation}
The color charge of quarks is invariant under the $SU(2)_{CS}$ transformation.
 Consequently the interaction of the color charge with the
chromoelectric field is $SU(2)_{CS}$-invariant. 
The interaction part of the QCD Lagrangian can be split
into the $SU(2)_{CS}$-invariant electric part and the 
$SU(2)_{CS}$-breaking magnetic part:
\begin{equation}
\overline{\psi}(x)   \gamma^{\mu}  T^a  \psi(x)  ~A^a_\mu = \psi^\dagger(x)T^a \psi(x) ~A^a_0 
  + \overline{\psi}(x)   \gamma^i   T^a   \psi(x) ~ A^a_i.
\end{equation}
In a given Lorentz frame interaction of
quarks with the electric part of the gluonic field is  chiral spin invariant like in electrodynamics.

The direct product of the $SU(2)_{CS}$ group with the flavor group
$SU(N_F)$ can be embedded into a $SU(2N_F)$ group. The latter group
contains the chiral group $SU(N_F)_R \times SU(N_F)_L \times U(1)_A$
as a subgroup.
The set of $(2N_F)^2-1$ generators of $SU(2N_F)$ is
\begin{align}
\{
(\tau^a \otimes {1}_D),
({1}_F \otimes \Sigma^n),
(\tau^a \otimes \Sigma^n)
\}
\end{align}
with $\tau$  being the flavor generators (with the flavor index $a$) and $n=1,2,3$ is the $SU(2)_{CS}$ index.\footnote{ For the extension of
the $SU(2)_{CS}$ and $SU(2N_F)$ algebra to Euclidean space see Refs. \cite{R1,G1}.}
The fundamental
vector of $SU(2N_F)$ at $N_F=2$ is

\begin{equation}
\Psi =\begin{pmatrix} u_{\textsc{R}} \\ u_{\textsc{L}}  \\ d_{\textsc{R}}  \\ d_{\textsc{L}} \end{pmatrix}. 
\label{fourcomp}
\end{equation}
\noindent
$SU(2N_F)$ is also a symmetry of the color charge and of the electric
interaction of quarks:
The  color charge and the electric part of the quark-gluon interaction have a 
$SU(2N_F)$
symmetry that is larger than 
the chiral symmetry   of  QCD as a whole. Of course, in order to discuss
the electric and magnetic components of the gauge field one must fix a reference frame. However, the invariant hadron mass is the rest frame energy.
Consequently, it is natural to choose the hadron rest frame to discuss
physics of the hadron mass generation. In a medium at high temperatures
the Lorentz invariance is broken and the preferred frame is the medium rest
frame.

The chiral spin and $SU(2N_F)$ symmetries of the electric part of the
Lagrangian are broken by the quark kinetic term, by the magnetic interaction,
by the $U(1)_A$ anomaly and by the quark condensate. They can be seen
as approximate symmetries in observables if, and only if the breaking effect is small, i.e., when the physics is dominated by the electric interaction. In QCD
the electric interaction is considered to be crucial for confinement of quarks,
because it is established in the case of the static quarks that confinement
is related to the linear interquark potential, which is connected to the area law of the Wilson loop. The area law, the linear potential and the formation
of the electric flux between the static quarks are very clearly seen on the
lattice, for a review and references see Ref. \cite{Bali}. Then it is natural to consider the $SU(2)_{CS}$ and $SU(2N_F)$
symmetries as symmetries of the confining interaction with ultrarelativistic
light quarks.

Given the $SU(2)_{CS}$ and $SU(2N_F)$ generators one can construct the
multiplets of both groups \cite{GP}. For $N_F=2$ these multiplets for
$J=1$ mesons are presented in the right panel of Fig. 1 \cite{GP}. For higher
spin mesons the multiplets look similar \cite{DGP}.

\section{Verification of the  $SU(2)_{CS}$ and $SU(4)$ symmetries.
$SU(2N_F) \times SU(2N_F)$ symmetry of confinement in QCD.
Implications for hadrons in vacuum}\label{sec:verification}

\begin{figure}
\centering 
\includegraphics[width=0.4\linewidth]{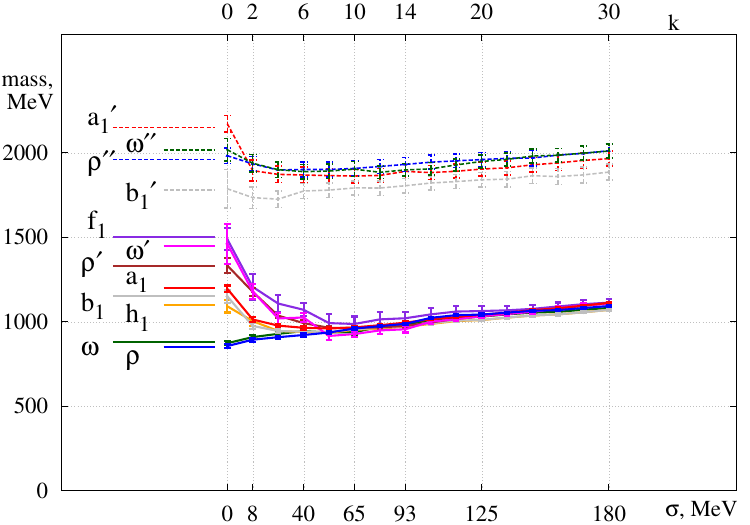}
\caption{$J=1$ isovector and isoscalar meson masses  as a function of the
truncation number $k$ where $k$ represents the amount of removed
lowest modes of the Dirac operator. $\sigma$ shows the energy gap in the Dirac spectrum. From Ref. \cite{D2}.} 
\label{den}
\end{figure}

The $SU(2)_{CS}$ and $SU(4)$ symmetry predictions, see the right panel of Fig. 1, have been
verified in Refs. \cite{D2,DGP}, where the low-mode truncation studies of $J=1,2$ mesons, both of isovector and isoscalar type, have been performed.
Similar results have been obtained with baryons \cite{DGP2}.
In Fig. \ref{den} we show the results for all $J=1$ mesons, both isovector and
isoscalar \cite{D2}. We observe a clear approximate degeneracy of
all $J=1$ mesons in the chirally symmetric world.
In particular all mesons from the 15-plet of $SU(4)$ are approximately
degenerate.

From these results we can infer a few important conclusions:
\medskip
\begin{itemize}%
\item While the quark condensate of the vacuum does contribute
 to some extent to the hadron mass, it would be incorrect to say
 that the hadron mass comes from the condensate, i.e., is a consequence of the spontaneous breaking of chiral symmetry. The role of the chiral symmetry
 breaking is to lift the $SU(4)$ degeneracy of the confining electric
 interaction.
 
\item The effects of magnetic interactions of quarks with gluonic
field are at least predominantly located in the near-zero modes of the Dirac operator.

\item The confining electric interaction is distributed in all
modes of the Dirac operator, not only in the near zero modes.
Consequently it would be incorrect to say that confinement is
an infrared phenomenon.

\item Confinement and spontaneous breaking of chiral symmetry
are not directly related phenomena, because while the effects of
confinement are distributed in all modes, the chiral symmetry
breaking is connected only to the near-zero modes. Still it is possible
that it is confinement which induces the spontaneous breaking of chiral
symmetry. However, the elimination of the chiral symmetry breaking,
e.g. in the medium at high temperatures, does not require deconfinement.
\end{itemize}

\medskip
In
reality the degeneracy seen in  Fig. \ref{den} represents a larger
symmetry. Indeed, both the 15-plet and the singlet of $SU(4)$ are
degenerate. The papers \cite{GP,Cohen} asked the question what symmetry is it? The latter
question was soon answered in an unpublished preprint and published
later  in Ref. \cite{G1}. It is a $SU(4) \times SU(4)$.
The irreducible 16-plet of $SU(4) \times SU(4)$
is a direct sum of the 15-plet and of the singlet of $SU(4)$.
The  physical reason for the emergence of the larger $SU(4) \times SU(4)$ symmetry
is the following. A confining electric flux tube binds a quark and an antiquark
and has two independent quark-gluon vertices. Each vertex has its own
$SU(4)$ symmetry. This simple physical picture can be rigorously
substantiated. With the local QCD Lagrangian this $SU(4) \times SU(4)$
symmetry cannot be seen. One needs the Hamiltonian that contains a bilocal
part.

 Consider the Minkowski QCD Hamiltonian in Coulomb gauge in the hadron rest frame \cite{Lee}:
\begin{equation}
H_{QCD} = H_{E,B}
 + \int d^3 x \Psi^\dag({\vec{x}}) 
[-i \vec{ \alpha} \cdot \vec{\nabla} ]  \Psi(\vec{x})
+ H_T + H_C,
\label{ham}
\end{equation}
\noindent
where the transverse (magnetic)  and instantaneous "Coulombic" parts to be:
\begin{equation}
H_T = -g \int d^3 x \, \Psi^\dag({\vec{x}}) \vec{\alpha} 
\cdot t^a \vec{A}^a(\vec{x}) \, \Psi(\vec{x}) \; , 
\end{equation}
\begin{equation} 
H_C = \frac{g^2}{2} \int  d^3 x \, d^3 y\, J^{-1} \ \rho^a(\vec{x})  
F^{ab}(\vec{x},\vec{y}) \, J \, \rho^b(\vec y) \; .
\label{coul}
\end{equation}
\noindent
Here $J$ is the  Faddeev-Popov determinant, $\rho^a(\vec{x})$ and $\rho^a(\vec{y})$
are the color-charge densities of quarks  and gluons at the space points  $\vec{x}$
and $\vec{y}$.
$F^{ab}(\vec{x},\vec{y})$ is a nonlocal
"Coulombic" kernel. 

The kinetic and transverse  parts of the Hamiltonian are
 chirally symmetric. The confining "Coulombic" part (\ref{coul})
carries the $SU(2N_F)$ symmetry, because the quark color charge density
operator is $SU(2N_F)$ symmetric. The gluonic part of the color charge density is trivially
 $SU(2N_F)$ invariant. 
However, both $\rho^a(\vec{x})$ and $\rho^b(\vec{y})$
 are independently $SU(2N_F)$ symmetric because the $SU(2N_F)$ transformations
 at  spatial points $\vec{x}$
and $\vec{y}$ can be completely independent, with different
rotations angles. 
This means that the confining "Coulombic" part of the Hamiltonian
 is  $SU(2N_F) \times SU(2N_F)$-symmetric, because it is invariant
 under the bilocal $SU(2N_F) \times SU(2N_F)$ transformation.
 
 \section{Hot QCD. Before and after RHIC}\label{sec:hotqcd}
 
 What happens with hadrons in a medium at small chemical potentials upon increasing temperature?
 At small temperatures we have a dilute hadron gas where hadrons
 practically do not interact and where the properties
 of hadrons are the same as in vacuum. This is so called hadron gas phase.
 This phase is characterized by spontaneous breaking of chiral
 symmetry and by confinement. Confinement is a difficult notion to define.
 In a dilute hadron gas one can use the most ancient definition that
 confinement means that asymptotically only the color-singlet hadrons
 exist, and there are no free quarks and gluons. Hagedorn argued long before the born of QCD
 that if the hadron spectrum possesses an exponential growth  of number of
 hadrons with excitation energy, then the hadron gas cannot be heated above
 the so called Hagedorn temperature \cite{Hag}, which is the limiting temperature
 in the system. With the advent of QCD Cabibbo and Parisi suggested that
  the Hagedorn temperature is actually a temperature of the
 phase transition from the hadron gas phase to the phase where the degrees
 of freedom would be deconfined (free) quarks and gluons \cite{Cab}. The
 asymptotic freedom requires that the strong coupling constant decreases
 upon increase of   temperature or chemical potential in the system \cite{Col}, consequently
 this could substantiate the transition from the hadron gas to the phase
 with deconfined quarks and gluons. When the coupling constant is small
 enough the perturbation theory can be used. Using the
 perturbation theory Shuryak has shown that at some temperature
 there should happen the Debye screening of the confining electric interaction,
 so the system would become deconfined with "free" quarks and
 gluons as degrees of freedom\footnote{Now it is clear that this temperature is very high, beyond the reach of the modern facilities.}. He called such a state of matter a quark-gluon plasma \cite{Shuryak}.
 These milestones gave the birth to the enormous experimental and
 theoretical effort to search and study the QGP.
 
 After the first experiments at RHIC on heavy ion collisions at large
 energies
 a quarter of century ago, where a high temperature arises in the fireball, it has
 become clear that the matter within the fireball is indeed very different
 from the hadron gas and the discovery of the quark-gluon plasma was announced. This matter is characterized by a high collectivity
 and a very small mean free path of the effective constituents; it resembles a liquid \cite{Heinz}. 
 It cannot   be a matter that consists of free  quarks and gluons
 that would form a gas, not a liquid. As a resolution of the paradox
 it was  suggested  that 
 this matter is a strongly interacting QGP, where the quark and gluon quasiparticles are deconfined but still strongly interact. Actually it is not
  clear what it means: "deconfined but strongly interacting quarks and gluons". There is no experimental evidence that after equilibration the degrees of freedom
 in the fireball at RHIC and LHC energies are the (quasi)quarks and
 (quasi)gluons.
 
 In parallel with experiments the high-temperature QCD matter in equilibrium was studied
 on the lattice. In particular it was established that there is no
 phase transition in the real world but instead a fast but smooth
 crossover takes place \cite{Aoki1}. The
pseudocritical temperature of the chiral symmetry restoration was determined
to be around 155 MeV \cite{Aoki2} and it was also found that the Polyakov loop \cite{Polyakov},
which is the order parameter for deconfinement in a pure glue theory or
in QCD with infinitely heavy quarks\footnote{The Polyakov loop is the order
parameter for the center symmetry of the pure glue theory. In QCD with light
quarks the center symmetry is explicitly broken by the quark loops so it is not any longer the order parameter for deconfinement. Still there was a hope
that the behavior of the Polyakov loop in full QCD would indicate the
"deconfinement" via its inflection point, which would be a remnant of
the first order deconfinement transition in the pure glue theory.
However, some possible "irregularity" in the Polyakov loop 
or in the behavior of the free energy 
in full QCD in the region of the fast chiral crossover should be
induced by the chiral crossover itself  which affects all observables at strong coupling. So it is misleading to interpret such "irregularity"
as deconfinement.}, demonstrates an inflection point at approximately
the same temperature (or slightly above) \footnote{Actually it turned out that
the latter statement was erroneous. The important issue is that the absolute value of the properly renormalized and normalized Polyakov loop is very small at the chiral restoration temperature, while the deconfinement in the pure glue theory at $T \sim 300$ MeV is
accompanied by a jump of the Polyakov loop from zero to the value
of the order 0.5-0.6.}. Consequently the community
 took the point that QCD undergoes a very fast crossover from the confined
hadron gas to a deconfined strongly interacting QGP around the temperature
155 MeV.

It will become clear from what follows that the conclusion about the
discovery of the QGP was premature and in reality another state of
the strongly interacting matter takes place at temperatures above
the chiral symmetry restoration temperature. 

Given the results on the artificial restoration of chiral symmetry discussed
in the previous section, the author suggested that above the chiral restoration temperature QCD should still be in the confining regime
that should be evidenced by the chiral spin symmetry which would naturally
emerge without any truncation \cite{G3}.

\section{Emergence of approximate chiral spin and $SU(4)$ symmetries
above chiral restoration crossover and its implications}\label{sec:emergence}

Symmetry properties of QCD can be studied via symmetries of
correlators calculated at a given temperature on the lattice. For  meson operators $O_\Gamma(t,x,y,z)=\bar{\psi}(t,x,y,z)\Gamma \frac{\boldsymbol{\tau}}{2}\psi(t,x,y,z)$ with 
$\Gamma\in\{1,\gamma_5,\gamma_\mu,\gamma_5\gamma_\mu,\sigma_{\mu\nu},\gamma_5\sigma_{\mu\nu}\}$, the Euclidean correlation functions,
\begin{equation}
C_\Gamma(t,x,y,z)=\langle O_\Gamma(t,x,y,z)\,O_\Gamma(0,\mathbf{0})^\dagger\rangle\;,
\end{equation}
carry the full dynamical information of all isovector excitations with $J=0,1$ \footnote{Note 
that at a finite temperature the
correlation functions   are automatically
calculated in the medium rest frame which is the preferred reference frame.}.
The spatial and temporal correlators in Euclidean space are defined as

\begin{equation}
C_\Gamma^s(z)=\sum_{x,y,t} C_\Gamma(t,x,y,z)\;,\label{eq:c_z}\\
\end{equation}

\begin{equation}
C_\Gamma^t(t)=\sum_{x,y,z}C_\Gamma(t,x,y,z)\;.
\label{eq:c_t}
\end{equation}
The temporal correlators reflect dynamics of the QCD Hamiltonian
since
$H$ translates states  in Euclidean time
\begin{equation}
|\psi(t+1;x,y,z)\rangle =\exp(-aH)|\psi(t;x,y,z)\rangle\;,\\
\label{Hz}
\end{equation}
where $a$ is the lattice spacing. Symmetry properties of the
$J=1$ operators which should be used in
temporal  
 correlators, are presented in Fig. 1. The $J=0$ correlators can address only chiral symmetries
because it is not possible to construct the multiplets of the chiral
spin and $SU(4)$ groups for $J=0$ operators, see for details Ref. \cite{G1}.
The spatial correlators are connected to the dynamics of the analogous operator $H_z$ translating states in 
$z$-direction
\begin{equation}
|\psi(t; x,y,z+1)\rangle = \exp(-aH_z)|\psi(t; x,y,z)\rangle\;.
\label{Hz}
\end{equation}
The chiral, $SU(2)_{CS}$ and $SU(4)$ multiplets relevant for the propagation along the spatial $z$ direction can be found in Ref. \cite{R1}.

\begin{figure}
  \centering
  \includegraphics[scale=0.4]{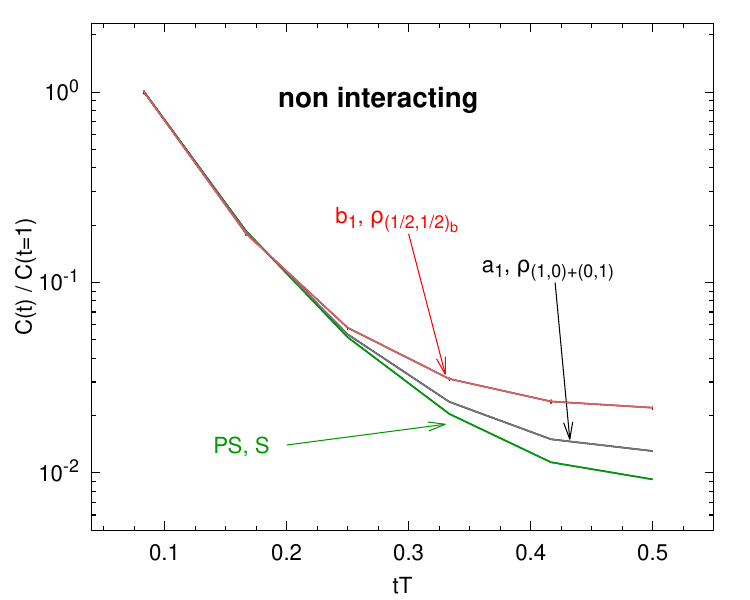} 
  \includegraphics[scale=0.4]{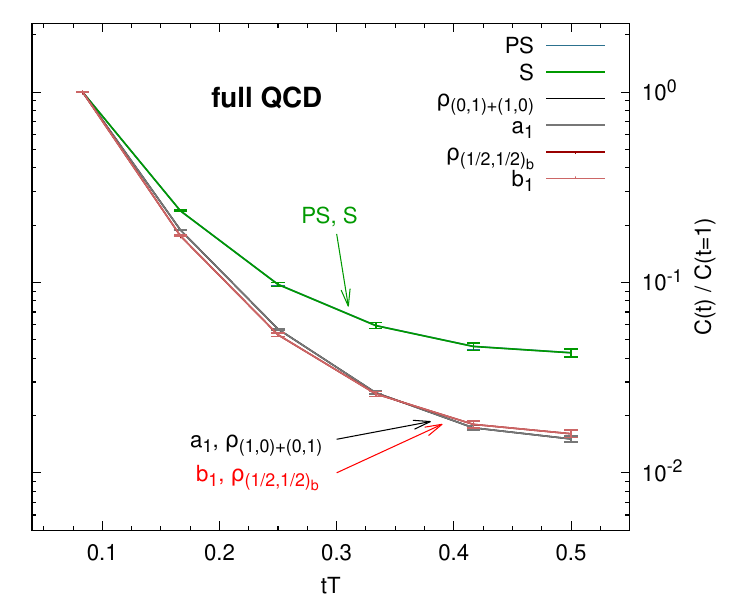} 
\caption{ Temporal correlation functions for $12 \times 48^3$
lattices. The l.h.s. shows correlators calculated with free
noninteracting quarks with manifest $U(1)_A$  and $SU(2)_L \times SU(2)_R$
symmetries. The r.h.s. presents full QCD results at a temperature 220 MeV,
which shows multiplets of all  $U(1)_A$, $SU(2)_L \times SU(2)_R$, $SU(2)_{CS}$  and $SU(4)$ groups. From Ref. \cite{R2}.
}
\label{tcorr}
\end{figure}

\medskip
We begin with the temporal correlators obtained in the two-flavor QCD with
a chirally symmetric Dirac operator \cite{R2}.
On the r.h.s of Fig.~\ref{tcorr} we show temporal correlators
(\ref{eq:c_t}) at $T=220$ MeV calculated 
at physical  $m_u = m_d$ masses. 
Emergence of the respective symmetries is signalled by  a degeneracy of
the correlators (\ref{eq:c_t}) calculated with operators
that are connected by the corresponding transformations.
The degeneracy of the $J=0$ isovector scalar (S) and pseudoscalar (P)
correlators reflects the approximately restored $U(1)_A$ symmetry.
The $U(1)_A$ breaking in QCD is induced by the quark condensate and by the
$U(1)_A$ anomaly. Above the chiral restoration crossover the
quark condensate vanishes. If there
is some small residual $U(1)_A$ breaking that is induced by the $U(1)_A$
anomaly, it is too small to be seen in the present data. The approximate
degeneracy of the isovector $J=1$ correlators $a_1, \rho_{(1,0)+(0,1)},
b_1, \rho_{(1/2,1/2)_b}$ indicates emerged $SU(2)_{CS}$ and
$SU(4)$ symmetries.

We can compare the correlators in full QCD with the correlators
calculated on the same lattice without the gluonic field, see
 Fig.~\ref{tcorr}, left panel. The latter correlators represent a free
 quark gas and would correspond to the quark-gluon plasma at a very
 high temperature, where the quark-gluon interaction can be neglected.
 In the free quark gas the  $U(1)_A$ and $SU(2)_R \times SU(2)_L$ chiral symmetries are unbroken which is clearly visible via exact degeneracies
 of the corresponding correlators. A qualitative difference between the pattern on the l.h.s and the pattern on the r.h.s of Fig.~\ref{tcorr} is
 appealing. In full QCD at $T=220$ MeV one observes not only expected
$U(1)_A$ and $SU(2)_R \times SU(2)_L$ chiral symmetries, but also approximate
emerged  $SU(2)_{CS}$ and
$SU(4)$ symmetries.

\begin{figure}
  \centering
  \includegraphics[scale=0.3]{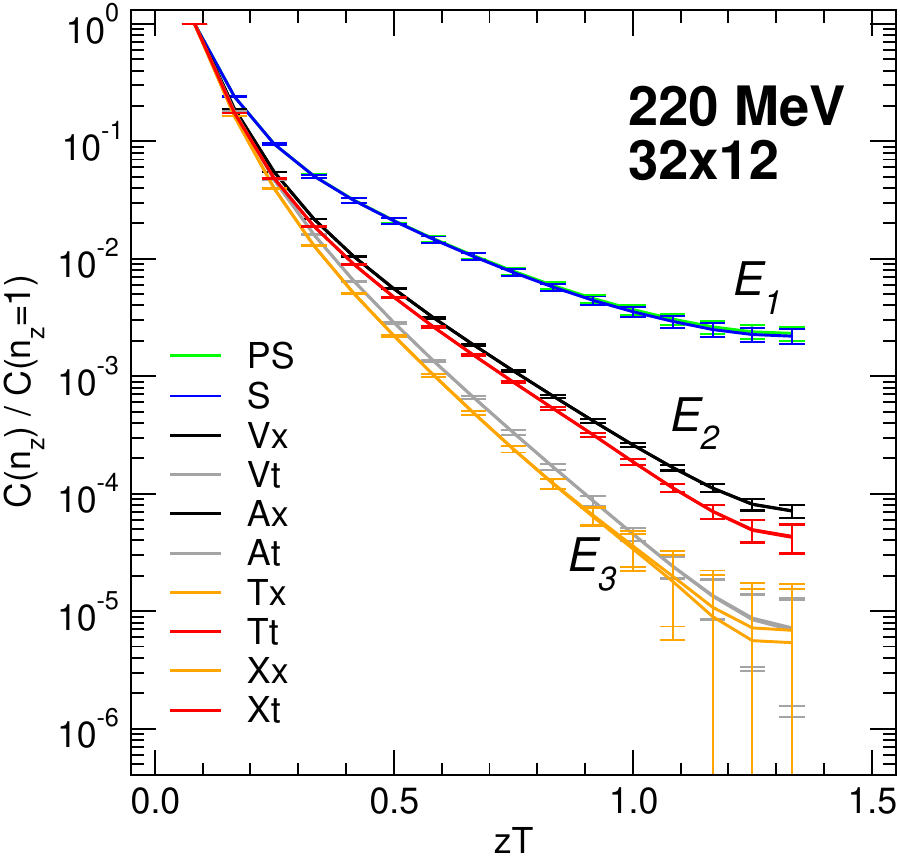} 
  \includegraphics[scale=0.3]{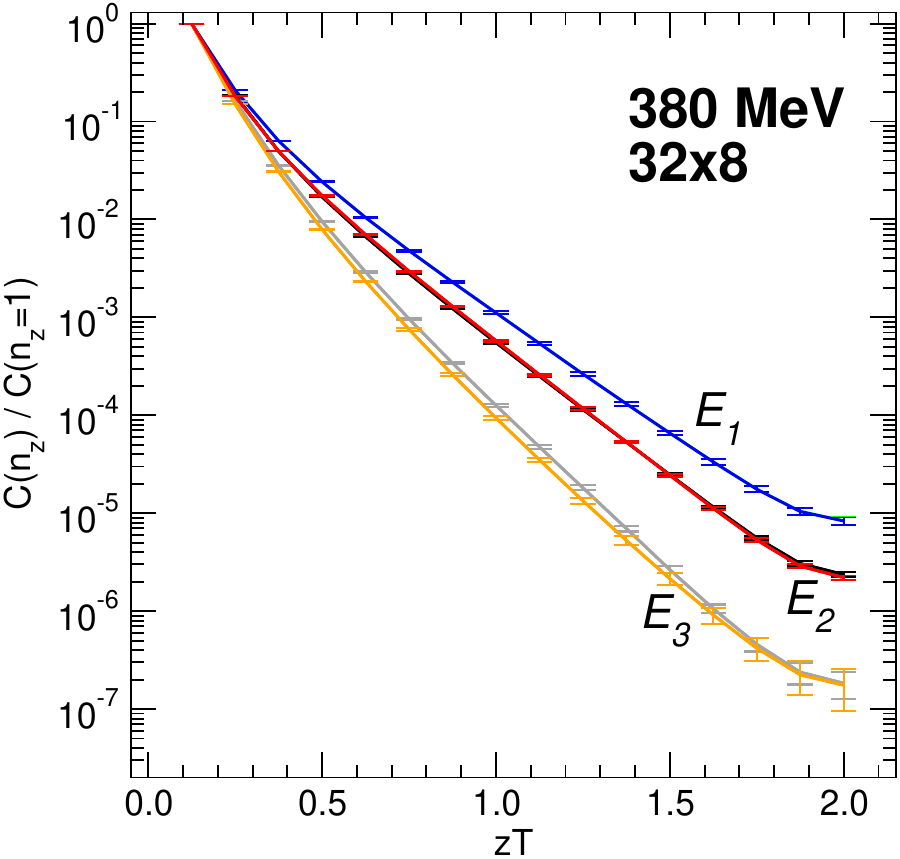} 
\caption{Spatial
correlation functions of all possible isovector $J=0,1$ bilinears.
  From
Ref. \cite{R1}.}
\label{spatial}
\end{figure}

\medskip
Now we will discuss observed symmetries of the spatial
correlators at different temperatures \cite{R3,R1}.
The $SU(2)_{CS}$ and $SU(4)$
transformation properties relevant to the spatial propagators
can be found in  Ref. \cite{R1}. 
In Fig. \ref{spatial} we show spatial correlators  (\ref{eq:c_z})
evaluated  with chirally symmetric domain wall
Dirac operator at physical quark masses within the $N_F=2$ QCD 
\cite{R1}. A complete set of all  isovector
 $J=0,1$ operators has been used.  We observe a distinct multiplet structure of the correlators.
This  structure  reflects symmetry properties of the
thermal partition function at the given temperature.

The multiplet $E_1$ consists of isovector scalar (S) and pseudoscalar (PS)
correlators. The degeneracy of S and PS correlators evidences restored,
at least approximately,
$U(1)_A$ symmetry. 
The multiplet $E_2$ contains four approximately
degenerate correlators obtained with $V_x, A_x, T_t, X_t$ 
 $J=1$ isovector operators. The $V_x$  and $A_x$  operators 
 are connected by the
  $SU(2)_L \times SU(2)_R$ transformation and 
their degeneracy evidences restored
$SU(2)_L \times SU(2)_R$ symmetry.
The $T_t$  and
 $X_t$  operators are connected by the
$U(1)_A$ transformation.  The operators $(A_x, T_t, X_t)$
form a triplet of the $SU(2)_{CS}$ group. The approximate degeneracy
of the corresponding correlators indicates emerged approximate $SU(2)_{CS}$ symmetry.
All four operators 
$(V_x, A_x, T_t, X_t)$ are connected by the $SU(4)$ transformation. 
The degeneracy of the corresponding correlators shows
emergent  approximate $SU(4)$ symmetry.
 The degeneracy of the normalized
$V_t, A_t, T_x, X_x$ correlators in the $E_3$ multiplet is consistent
with chiral symmetry alone: it cannot be used as an indicator
of emerged chiral spin symmetry. 

We observe approximate emerged $SU(2)_{CS}$ and $SU(4)$ symmetry up
to temperatures of about $\sim 500$ MeV. At higher temperatures
the distinct multiplet $E_2$ disappears. This would
naturally happen because of the Debye screening of the electric confining interaction.
The presented  results on symmetries of temporal and spatial
correlators with $N_F=2$ QCD have been
confirmed in $N_F=2+1+1$ QCD \cite{Chiu}.

\medskip
We conclude this section with a summary of the observations
made for QCD.
The QCD thermal partition function above 
the chiral crossover has not only chiral symmetries
but is approximately symmetric with respect to
$SU(2)_{CS}$ chiral spin group and its flavor extension $SU(4)$.
  The approximate chiral spin symmetry
 can emerge  only
when the confining electric interaction strongly dominates over
the magnetic interaction and over the quark kinetic
term. This symmetry is characteristic of the quark-antiquark
systems with the chirally symmetric quarks bound into the color-singlets by the chromoelectric
field.
The emergent $SU(2)_{CS}$ and $SU(4)$ symmetries 
suggest that the propagating physical degrees of freedom at these temperatures
are chirally symmetric quarks bound into the color-singlets by the
confining chromoelectric field.
The stringy fluid regime arises above $T_{ch}$ and extends to
roughly $3 T_{ch}$.
Above these temperatures the chiral spin symmetry 
disappears because the confining electric field gets screened
and one observes  a quark-gluon plasma.

From the symmetry data we can conclude that the deconfinement crossover
is very smooth. One might assume that in QCD it is a very broad crossover
around the deconfinement phase transition of the pure glue theory at a temperature $T_d \sim 300$ MeV \cite{CG1,CG2}. This assumption is supported by recent results on
center vortices percolation in full QCD at different temperatures \cite{Allton}. 

\section{Three regimes/phases of QCD and their $N_c$ scaling}\label{sec:threeregimes}

The symmetry studies suggest a three-regimes structure of the
QCD phase diagram at small chemical potential. These regimes
are separated by smooth crossovers and differ in symmetries and degrees
of freedom. Below the chiral restoration temperature $T_{ch} \sim 155$ MeV
the QCD matter is a dilute hadron gas with confinement and spontaneously broken chiral symmetry. Hadrons are very well separated and keep properties
of hadrons in vacuum. Above the chiral restoration crossover and below
a very smooth deconfinement crossover centered around $T_d \sim 300$ MeV,
one finds the stringy fluid regime where chiral symmetry is restored but
 the propagating in time degrees of freedom are still the color-singlet systems. This regime is
characterized by approximate $SU(2)_{CS}$ and $SU(4)$ symmetries. The regime
is still with confinement. Around $T_d$ one has a very smooth crossover
to QGP. The QGP is characterized by chiral symmetry, by the absence
of the chiral spin symmetry, where the effective degrees of freedom are deconfined, i.e. the
quark and gluon quasiparticles. There are also other lattice evidences,
not related to symmetries, which support the three-regimes picture, see
a review \cite{G1}.

\begin{figure}
  \centering
  \includegraphics[scale=0.3]{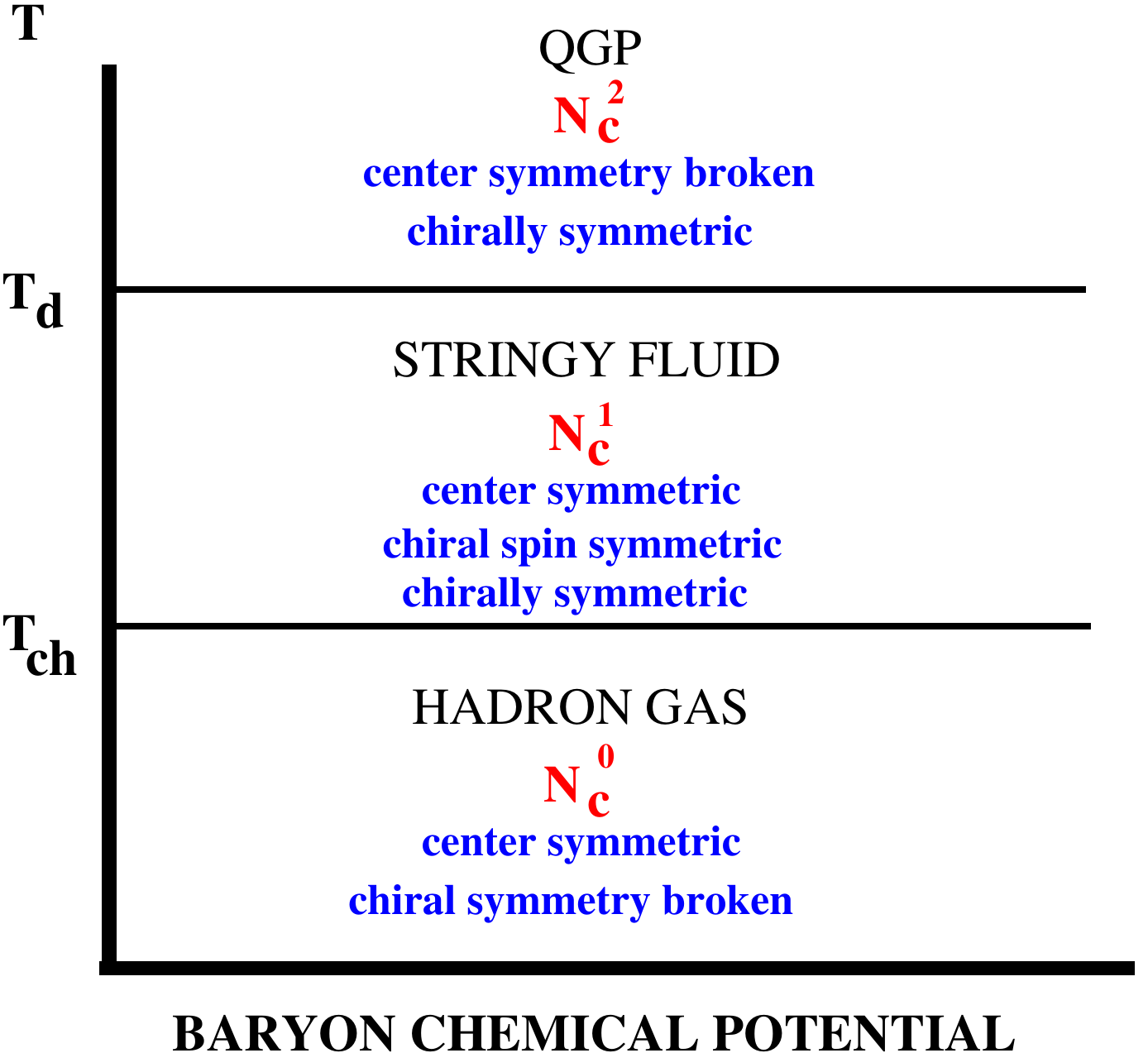}  
\caption{Three regimes/phases of QCD at small chemical potential
with their symmetries and $N_c$ scaling. With a sufficiently large $N_c$ 
the quark loops are suppressed  and the center symmetry is a  relevant symmetry; the order parameter for deconfinement/confinement  is then the Polyakov loop. In the large $N_c$
limit one expects that the stringy fluid regime becomes a distinct phase
separated from the hadron gas and the QGP by phase transitions.}
\label{three}
\end{figure}

It has been suggested in ref. \cite{CG1} that these three regimes are
 characterized by different scaling of the energy density $\epsilon$, pressure $P$
and entropy density $s$ with $N_c$, where $N_c$ is the number of colors in QCD,
see Fig. \ref{three}:
\begin{equation}
    \epsilon_{\rm HG} \sim N_c^0 \; \; , \; \; P_{\rm HG} \sim N_c^0 \; , \; s_{\rm HG} \sim N_c^0 \; , \label{Eq:had scale}
\end{equation}
 
\begin{equation}
    \epsilon_{\rm str} \sim N_c^1 \; \; , \; \; P_{\rm str} \sim N_c^1 \;  ,  \; s_{\rm str} \sim N_c^1 \;  ,
     \label{Eq:IntScal}
\end{equation}

\begin{equation}
    \epsilon_{\rm QGP} \sim N_c^2 \; \; , \; \; P_{\rm QGP} \sim N_c^2 \;  ,  \; s_{\rm QGP} \sim N_c^2 \;  .
     \label{Eq:IntScal}
\end{equation}

The $N_c^0$ scaling of the thermodynamic quantities 
in the hadron gas  will be discussed below. The $N_c^2$ scaling within the QGP is for the following reason. 
In the deconfined phase there are
$N_c^2-1$ independent gluon quasiparticles and only $N_c$ quark quasiparticles. Consequently  it is the gluon quasiparticles which dominate
the thermodynamic quantities in the equilibrium.
The $N_c^1$ scaling within the new regime, the stringy fluid, 
was suggested in Ref. \cite{CG1}. The physical reason for the latter scaling
within the confined but chirally symmetric matter  was clarified  in Ref. \cite{G4} and will be discussed below.

The different $N_c$ scaling of the three regimes has important
consequence for the structure of the phase diagram. When $N_c$
is sufficiently (infinitely) large one expects that the energy density experiences infinite jumps at $T_{ch}$ and $T_{d}$ and instead of three regimes with smooth crossovers in the real world $N_c=3$ one obtains three
distinct phases with first order phase transitions between them \cite{CG1}.
At large $N_c$ the quark loops get suppressed and the gluodynamics becomes
effectively quenched. In this case the center symmetry of the pure glue theory gets relevant
and one can unambiguously define confinement-deconfinement phase
transition via the Polyakov loop which is the order parameter for
the center symmetry. The temperature of the deconfinement phase transition in the pure glue theory is known to be practically independent of $N_c$ \cite{Lucini}.

The three regimes/phases picture of Fig. \ref{three} should be a
subject for further lattice studies. The equation of state on the lattice
at $N_c=3$ has been explored \cite{Bor,hot,bpw}. One should repeat  measurements
of the thermodynamic quantities at $N_c=5,7$, which is rather straightforward,
and establish the $N_c$ scalings of the equation of state
at different temperatures. This would be a complementary and important lattice information about the QCD phase structure.

\subsection{Origin of the $N_c^0$  scaling in hadron gas}\label{subsec:gas}

At $N_c=3$ and vanishing baryon chemical potential the thermodynamics
of the hadron gas
is dominated by the light mesons but baryons also contribute. If the number of
colors in QCD, $N_c$, increases, baryons decouple from thermodynamics
because they are much heavier than mesons \footnote{Meson masses scale
as $N_c^0$, while baryon masses as $N_c^1$ \cite{Hooft,Witten}.}. Hence
we approximate the hadron gas as a meson gas. Mesons in a dilute gas
interact weakly; their interaction vanishes as $N_c^{-1}$ \cite{Witten}. Consequently we can approximate the
meson gas via the Bose-Einstein distribution for the number density $n_k$ of species  $k$  and their energy density $\epsilon_k$:
\begin{equation}
  n_k(T)  = (2S_k+1)(2I_k+1) \int \frac{{\rm d}^3 p}{(2 \pi)^3} \, \frac{1}{e^{\sqrt{p^2 + m_k^2}/T} - 1}  \; ,
\end{equation}
\begin{equation}
 \epsilon_k(T)  = (2S_k+1)(2I_k+1) \int \frac{{\rm d}^3 p}{(2 \pi)^3} \, \frac{\sqrt{p^2 + m_k^2}}{e^{\sqrt{p^2 + m_k^2}/T} - 1}  \; .
\end{equation}
Since $ m_k \sim N_c^0$, one gets that $n_k(T) \sim N_c^0$  and $\epsilon_k(T)
\sim N_c^0$ in the meson gas. Similar derivations can be found for  the pressure
and entropy density in Ref. \cite{CG1}.

The ideal hadron gas predictions  reproduce well the lattice results
for thermodynamics at $T < T_{ch}$ and radically deviate from the lattice data
at  $T > T_{ch}$ \cite{Bor,hot,bpw}. This tells that the nature of the chirally symmetric 
regime with confinement is very far from a dilute noninteracting gas of the color-singlet hadrons. We will clarify a possible microscopic picture
for the stringy fluid below.

\subsection{Fluctuations of conserved charges as evidence for  stringy fluid}\label{subsec:fluctuations}

The success of the hadron gas picture below $T_{ch}$ indicates that at these temperatures
the hadron structure is not yet resolved and its internal degrees of freedom
are frozen. This is the reason for the $N_c^0$ scaling of the thermodynamic
observables in the hadron gas regime.
Once the density of hadrons increases and they start to overlap the internal hadron structure gets relevant. Here we will concentrate on an observable
that clearly distinguishes the hadron gas from the stringy fluid, which can be measured experimentally  and has been calculated on the lattice.
This observable is fluctuations of conserved charges \cite{Asakawa}.

Consider charges associated with the net number of u,d,s quarks.
\begin{equation}    
N_q \equiv \int d^3 x ~n_q(x) \; \;\;  {\rm with}
\; \; \;n_q(x) = \bar q(x) \gamma^0 q(x), \;\; \; q=u,d,s
\label{def}
\end{equation}
Each quark can be in one of the  $N_c$ color states and contraction with respect
to the color of quarks is assumed. This means that the  conserved
flavor charges  $N_q$ scale as $N_c^1$ \cite{CG2}.

In experimental measurements as well as on the lattice one usually (but not
always)
deals with the  baryon charge $B$, electric charge $Q$ as well as strangeness $S$,
which are conserved in strong interactions. All these quantities can be related
with the net number of $u,d,s$ quarks in the system, see Ref. \cite{CG2}.
The latter quantities are more convenient as the $N_c$ scaling is very
clear for the quark number while it is obscured for $B,Q,S$. 
 While in the hadron
gas the hadron structure is frozen and all quantities scale as $N_c^0$,
in the stringy fluid the quark content is of relevance. 

Consider a dense strongly interacting stringy fluid matter.
At zero baryon chemical potential, vanishing isospin
and strangeness chemical potentials, the number, e.g.,
of $u$-quarks and $u$-antiquarks is the same, so the expectation value of the conserved charges vanishes,
$<N_u>=0, <N_d>=0, <N_s>=0$.  However, the
fluctuations of the conserved charges do not vanish and can be measured.
Indeed, the variance of $N_q$ is
\begin{equation}
<\delta N_q^2> = <(N_q -<N_q>)^2> = \int d^3 x_1 d^3 x_2 
<\delta n_q(x_1)\delta n_q(x_2)>,
\label{var}
\end{equation}
where $\delta n_q(x) = n_q(x) - <n_q>$.
 Since $N_q$ scales as $N_c^1$ one obtains that the variance
of the quark conserved charges scales as $N_c^2$. The linear measure of
 fluctuations, the standard deviation, is square root of variance, scales as $N_c^1$:
\begin{equation}
 \sigma_{N_q} = \sqrt {<\delta N_q^2>} \sim N_c^1.
\label{sd}
\end{equation}
Thus the fluctuations of conserved quark charges scale as $N_c^1$
in the stringy fluid \cite{CG2}.

\begin{figure}[h]
\centering
 \includegraphics[width=0.3 \textwidth]{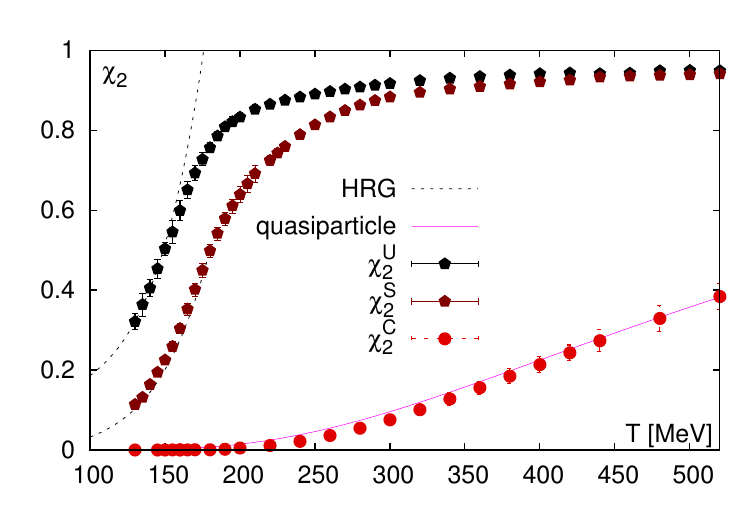}
  \includegraphics[width=0.3 \textwidth]{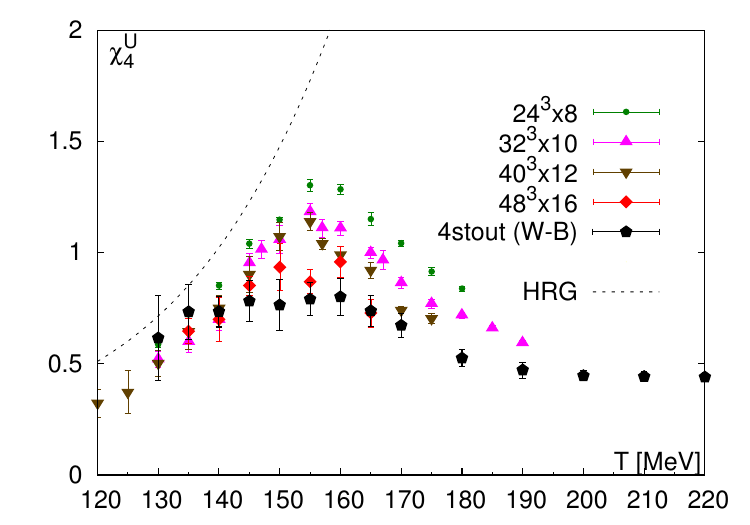} 
  \caption{Left panel: Fluctuations of conserved net $u$ ($\chi_2^U= \chi_{2,0,0}^{u,d,s}$) and strange ($\chi_2^S= \chi_{0,0,2}^{u,d,s}$) quark numbers in 2+1 QCD at physical quark masses. $\chi_2^C$
    is irrelevant to our discussion and can be ignored. 
    Right panel: Cumulant $\chi_4^U= \chi_{4,0,0}^{u,d,s}$  in 2+1 QCD at physical quark masses. From. Ref. \cite{Bel}.  }
    \label{fluc}
\end{figure}

The expectation value of the conserved quark number of a given flavor in  volume $V$ at temperature $T$ can be
obtained from the grand canonical partition function as

\begin{equation}
<N_i> = \frac{T \partial \left[log Z(T,V,\mu_u,\mu_d,...)\right]}{\partial \mu_i}.
\end{equation}
The fluctuations of conserved charges  can be calculated as a derivative
of these charges

\begin{equation}
\frac {\partial <N_i>}{\partial  \mu_j} = \frac{T \partial^2 \left[log Z(T,V,\mu_u,\mu_d,...)\right]}{\partial \mu_j \partial \mu_i}.
\end{equation}
The fluctuations and correlations of conserved charges can be expressed in terms
of different cumulants. 
\begin{equation}
\chi_{i,j,k}^{u,d,s} = \frac{T \partial^{i+j+k} \left( P/T^4\right)  }{(\partial\mu_u)^i (\partial \mu_d)^j  (\partial \mu_s)^k }.
\end{equation}
In Fig. \ref{fluc}  we show typical results for fluctuations of quark numbers 
of $u,d,s$ quarks taken from Ref. \cite{Bel} and their comparison with the
hadron gas model. We see
that the fluctuations of the $u,d,s$ quark numbers deviate from
the hadron gas (HRG) just at the chiral restoration temperature 155 MeV. 
Since the scaling of the fluctuations of conserved charges in the hadron gas
is $ N_c^0$ and it is $ N_c^1$  in the stringy fluid, one observes a clear change of the scaling $N_c^0 \rightarrow N_c^1$ just across
the chiral transition.
 The fluctuations and correlations of the conserved charges measured on the lattice
 indicate a transition from the hadron gas to a regime with the scaling $N_c^1$.
This statement entirely follows from the definition of the conserved quark charges (\ref{def}).  It could
be verified on the lattice via simulations at $N_c > 3$.

\subsection{Origin of the  $N_c^1$ scaling of  energy density in   stringy fluid}\label{subsec:stringy}

Above $T_{ch}$ in the chirally symmetric but confined stringy fluid the situation is qualitatively
different as compared to the hadron gas, because the color-singlet quark-antiquark systems are densely packed
and interact strongly. In this case the Bose-Einstein distribution is not
applicable.

What is the $N_c$ scaling of the excitation energy $E_k$ in the color-singlet
quark-antiquark system above $T_{ch}$? The Bethe-Salpeter equation
that describes these excitations is $N_c$-independent \cite{G4}.
Consequently the excitation energy of the bound color-singlet quark-antiquark
system scales as $N_c^0$ not only in the vacuum or hadron gas, but also
within the confining stringy fluid. If one describes these color-singlet
objects as strings \cite{ffhl}, then their energy is also of the order
$N_c^0$, because the string tension constant does not depend on $N_c$.

In order to evaluate the energy density one needs also a substitute
of the number density in a dense strongly interacting medium.
At vanishing chemical potentials all expectation values 
of the bilinear color-singlet operators $\bar q \Gamma_k q$ (where $\Gamma_k$ includes $\gamma$- and flavor-matrices) with not vacuum quantum numbers automatically vanish,
$<\bar q \Gamma_k q>=0$.
Consequently they can not be used as a substitute of the meson number density. However, fluctuations (variance) of these quantities
do not vanish and can be used to evaluate the energy density above $T_{ch}$
in a dense medium.\footnote{In order to understand this we remind the
reader the high-school physics of ideal gases. The mean value of velocity of molecules  vanishes, $ < \vec v > =0.$
But obviously the molecule's velocity is not zero, but is actually large.
We evaluate the typical velocity via fluctuations
$ \sqrt {< {\vec v}^2 >} = \sqrt {< {\vec v}\cdot {\vec v} >}$.}
So we estimate the energy density in a dense medium as a product  of the  
 linear measure of the fluctuations   of such color-singlet quark-antiquark pairs, the standard deviation,
 \begin{equation}
 \sigma_k =\sqrt{\int d^3 x_1 d^3 x_2 <\bar q(x_1) \Gamma_k q(x_1) \bar q(x_2) \Gamma_k q(x_2)>} ,
 \label{fl}
\end{equation}
 and the energy  $E_k$ of each pair:
\begin{equation} 
\sum_k \sigma_k E_k  
\label{de}.
\end{equation}
Like for conserved charges, we have
 $\sigma_k \sim N_c^1$.
Consequently one obtains  the $N_c^1$ scaling of the energy density in the
stringy fluid, 
$\epsilon_{\rm str} \sim N_c^1$ \cite{G4}.
 Similarly one can derive
the same scaling of other thermodynamic quantities. This scaling prediction
should be checked on the lattice via simulations at $N_c > 3$.

\section{Possible microscopic picture of  stringy fluid}\label{sec:model}

One can employ known manifestly confining and chirally
symmetric model  \cite{LeYaouanc:1983huv,LeYaouanc:1984ntu,Adler:1984ri,Kocic:1985uq,Bicudo:1989sh,Bicudo:1989si,Llanes-Estrada:1999nat,Wagenbrunn:2007ie}
in order to get insight into the
structure of the chirally symmetric stringy fluid with confinement.
This model  in the vacuum was solved long ago and the chiral symmetry breaking
was obtained as a consequence of the confining interaction \cite{Adler:1984ri}.
The model has been applied at finite temperatures, the chiral symmetry
restoration phase transition has been observed \cite{reinhardt,gnw1} and the
properties of the color-singlet chirally symmetric quark-antiquark bound 
 states have been studied \cite{gnw2}. Here we will rely on  results
 obtained in Refs. \cite{gnw1,gnw2}.

The model Hamiltonian is an approximation to the QCD Hamiltonian in the Coulomb
gauge. The gluonic part retains only  the instantaneous confining linear
potential between the color charge densities of quarks at the spatial
points $x$ and $y$\footnote{The  linear confining Coulomb potential is the
apriori input into the model and is kept  to be independent of temperature,
which is a qualitative assumption that can be justified at $T < T_d$ at large $N_c$, where
the quark loops are absent.}
\begin{equation}
H=\int d^3x\;\psi^\dagger(\vec{x},t)\left(-i\vec{\alpha}\cdot
{\vec \nabla}+\beta m_q\right)\psi(\vec{x},t)
 + \frac{1}{2} \int d^3x\; d^3y\;\rho^a(\vec{x})K_{ab}(|\vec{x}-\vec{y}|)\rho^b(\vec{y}),
\label{GNJL}
\end{equation}

\begin{equation}
K_{ab}(|\vec{x}-\vec{y}|)=\delta_{ab}V_0(|\vec{x}-\vec{y}|).
\label{Kab}
\end{equation}
The quark kinetic part is chirally symmetric while the confining  part is invariant under larger symmetry groups: $SU(2)_{CS}$, $SU(4)$, and $SU(4) \times SU(4)$ for two degenerate flavors.
A standard approach to solving the model implies a rainbow approximation for the dressed quark Green's function and a ladder approximation for the quark-antiquark Bethe--Salpeter equation. Such approximation is well justified in the large-$N_c$ limit. So while the model has been solved at $N_c=3$
it keeps the spirit of the large $N_c$ theory.  
A linearly rising confinement potential is
\begin{equation}
V_{\rm conf}(r)= C_F V_0(r)=\sigma r,
\label{Vlin}
\end{equation}
with $C_F$ being the color Casimir factor and $\sigma$ represents the fundamental "Coulomb string tension".  
An  attractive feature of the considered model  is its direct analogy 
with the 't~Hooft model for QCD in 1+1 dimensions in the large-$N_c$ limit \cite{tHooft:1974pnl} that was extensively studied in the axial (Coulomb) gauge in Refs.~\cite{Bars:1977ud,Kalashnikova:2001df,Glozman:2012ev}. In the 't ~Hooft model the appearance of a linearly rising potential between quarks is a consequence of the form of the two-dimensional gluon propagator in the Coulomb gauge. 

\begin{figure}[t!]
\centering
\includegraphics[width=0.35\columnwidth]{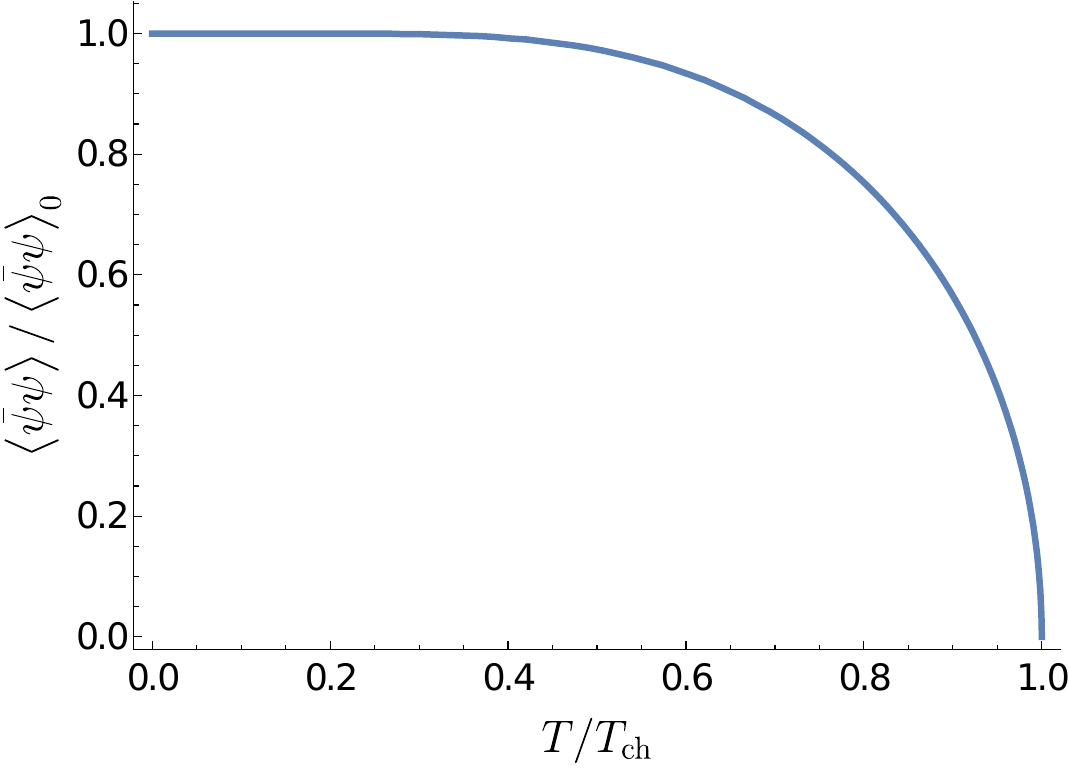}
\includegraphics[width=0.35\columnwidth]{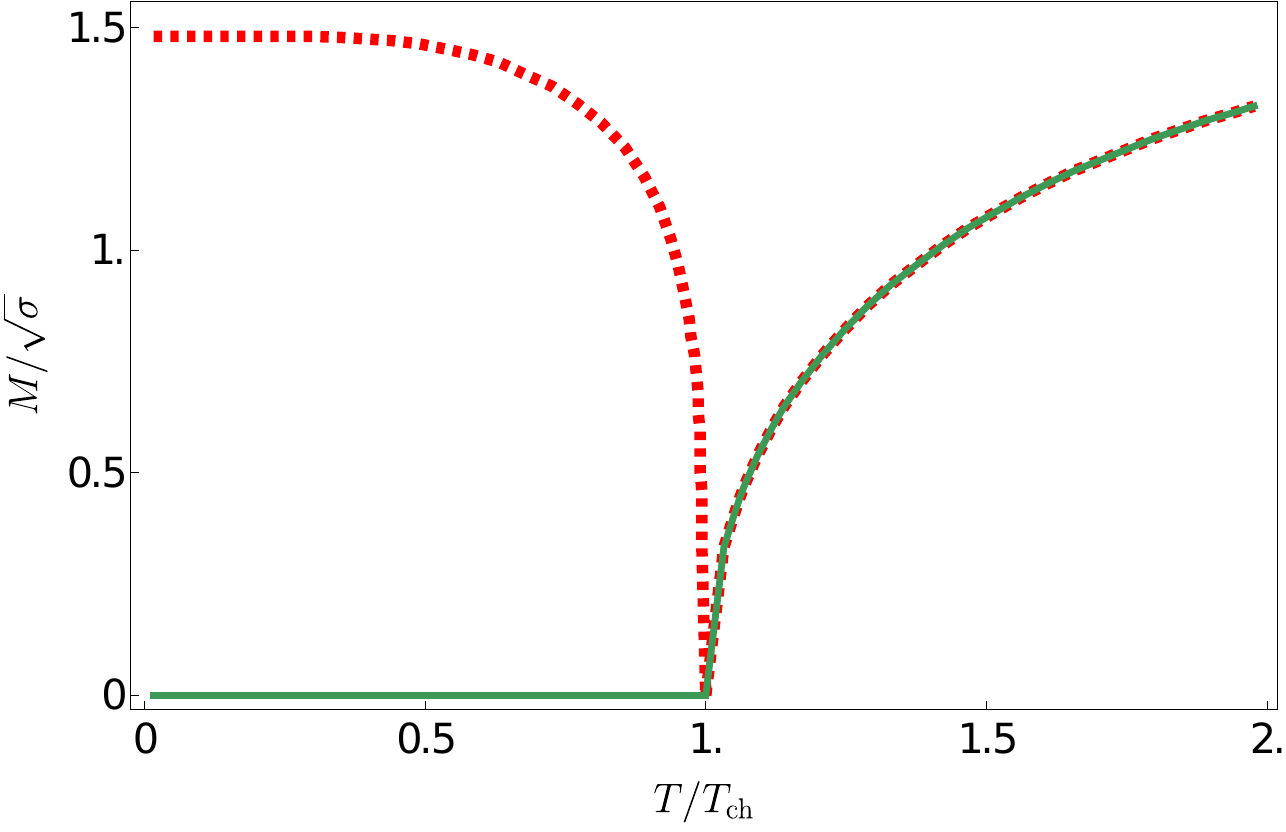}
\caption{Left panel:Temperature dependence of the chiral condensate  normalized to its maximum value reached at $T=0$. Right panel:
The masses of the pseudoscalar (green solid line) and
scalar (red dotted line) mesons in the units of $\sqrt{\sigma}$ as a function of $T/T_{ch}$. From Ref. \cite{gnw2}.}
\label{fig:scalarpseudoscalar}
\end{figure}

With a confining  interaction, the  chirally-symmetric vacuum is unstable. This instability can be studied with the help of the Bogoliubov-Valatin transformation of the quark field that leads to the gap equation. The
solution of the gap equation demonstrates the chiral symmetry breaking in the vacuum with the quark condensate 
$
 <{\bar{\psi}\psi}>_{T=0}\approx-(0.23\sqrt{\sigma})^3
$ \cite{Adler:1984ri}.
 The finite temperature gap equation was derived and solved in Ref.
\cite{gnw1}. At the critical temperature $T_{ch}\approx 0.084\sqrt{\sigma}$
 the chiral
symmetry gets restored and the chiral condensate vanishes, see the left panel of Fig. \ref{fig:scalarpseudoscalar},
With the phenomenological value for the chiral condensate $<{\bar{\psi}\psi}>_0=-(250~\mbox{MeV})^3$, it predicts
$T_{ch}\approx 90$ MeV,
which should be compared with the lattice chiral phase transition temperature, $T_{ch} \simeq 130$ MeV \cite{kkk}.
The  reason of the chiral restoration in the confining regime is Pauli
blocking of the quark and antiquark levels, required for  a non-vanishing
quark condensate, by the thermal excitation of quarks and antiquarks.

\begin{figure}[t!]
\centering
\includegraphics[width=0.65\textwidth]{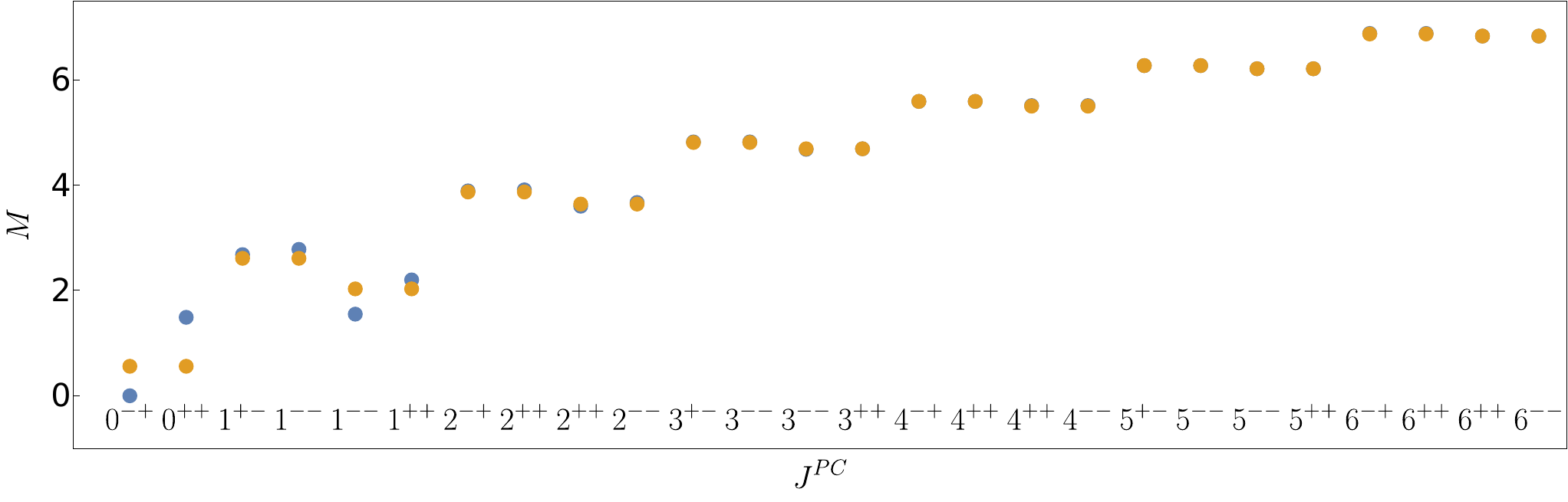}
\caption{Masses (in the units of $\sqrt{\sigma}$) of the  lowest radial states at $T=0$ (blue) and $T=1.1T_{ch}$ (yellow). From Ref. \cite{gnw2}.}
\label{fig:grstates}
\end{figure}

Given the single quark Green function at a finite temperature, obtained from the gap
equation, one can address the quark-antiquark bound states via
solution of the Bethe-Salpeter equation \cite{gnw2}. The Bethe-Salpeter
equation was solved in all possible quark-antiquark channels with
all possible $I, J^{PC}$ and all possible chiral representations.
The chiral symmetry in the meson spectrum above $T_{ch}$ is seen from the
degeneracy of the lowest pseudoscalar and scalar states shown in the right panel of Fig. \ref{fig:scalarpseudoscalar} as well as from the emergence of
all possible chiral multiplets for the states with different $J$, shown in Fig.
\ref{fig:grstates}.
However, the spectrum with $J>0$ demonstrates above  $T_{ch}$ not only
chiral symmetry, but also approximate $SU(4) \times SU(4)$ symmetry
of confinement. The latter symmetry requires a degeneracy of all possible
states at a given $J > 0$. We observe such approximate degeneracy which
improves with increased $J$.

The solution of the Bethe-Salpeter equation delivers not only
the excitation spectrum of the quark-antiquark systems, but also
their "wave functions". The wave functions of the systems with
different quantum numbers are discussed in detail in Ref. \cite{gnw2}.
\begin{figure}[t!]
\centering
\includegraphics[width=0.45\textwidth]{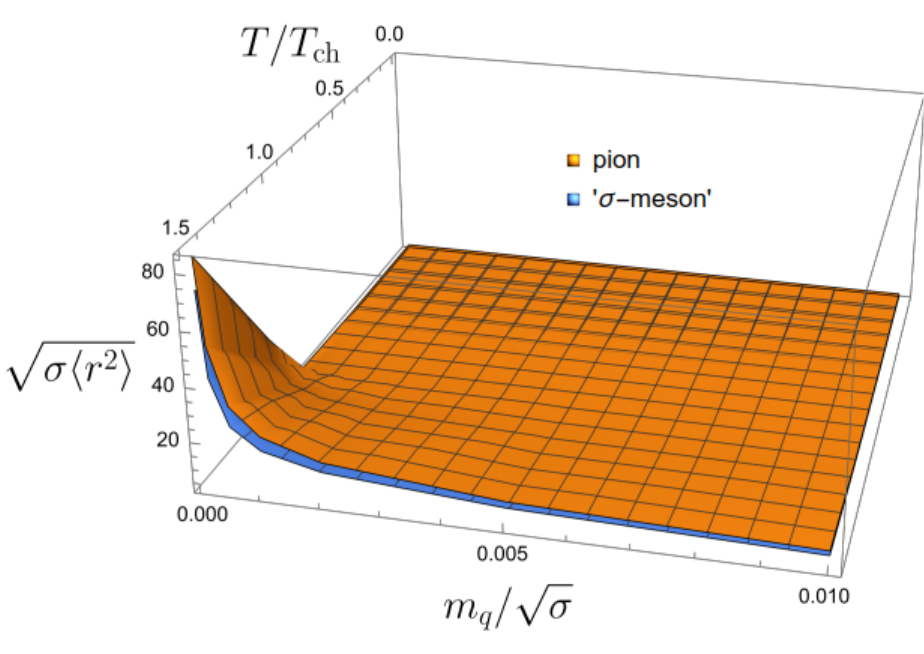}
\caption{3D plot for the r.m.s. radius of the pion (yellow) and lowest
scalar meson (blue) as function of the quark mass and temperature.
From Ref. \cite{gnw2}.}
\label{fig:radius3D}
\end{figure}
At $T=0$ and all temperatures below $T_{ch}$ the meson wave functions of
low spin mesons are localized in a small space volume. This is a result
of the confining interaction between quarks which acquire a dynamical mass
due to spontaneous breaking of chiral symmetry. At the chiral restoration
temperature the thermal excitations of quarks and antiquarks block the levels required for the existence of a non-vanishing quark condensate: the chiral symmetry gets restored. This Pauli blocking of the quark levels with small
momenta weakens effectively  the linear confining potential and the color-singlet quark-antiquark system gets delocalized. This delocalization
 at realistic quark masses increases the root-mean-square radius of the low-spin mesons several times as compared to their size in vacuum, while in the chiral
limit these systems become infinitely large, see Fig. \ref{fig:radius3D}.
The delocalization of the color-singlet quark-antiquark systems does
not mean that the confinement property is lost. Indeed, the confinement
potential $V_{conf}$ is assumed to be the same both below and above $T_{ch}$.
Consequently, only the color-singlet states are allowed and their spectrum is discrete; there are no single
quark poles in the complex energy plain.

This large swelling of the low-spin mesons above $T_{ch}$ has  remarkable phenomenological
implications: the stringy fluid matter is a system of huge overlapping
low-spin color-singlet quark-antiquark objects. This automatically implies
that the system is highly collective with a very small mean-free path
of the color-singlet constituents. The latter properties were experimentally observed at RHIC and LHC to be the most important features of the hot
QCD matter above the hadron gas phase. 

Of course, the swelling of the low-spin mesons above $T_{ch}$ is a qualitative
result obtained within the model where many possible QCD effects are missing. 
Consequently, one cannot expect that the specific values for the size of the pion-like state 
shown in Fig. \ref{fig:radius3D}, should be valid in QCD.
It is important to mention that the very fact of the pion swelling at
$T_{ch}$ can be proven to be true in QCD in a model-independent way,
though a specific value for the radius of the pion-like excitation above $T_{ch}$ should
be clarified in future lattice studies \cite{YaGlozman:2026ywk}.

\section{QCD phase diagram in the chiral limit}\label{sec:limit}

In the chiral limit the smooth chiral symmetry restoration crossover
around $T_{ch} \sim 155$ MeV becomes a second order chiral restoration
phase transition at the temperature  $T_{ch} \sim 130$ MeV \cite{kkk}.
Consequently, in the chiral limit the stringy fluid is the phase
of QCD above the hadron gas phase. At some high temperatures this
phase should smoothly transform into what is known as the quark-gluon plasma
with deconfined quark and gluon quasiparticles as degrees of freedom.
There is only one phase transition, if we keep $N_c=3$, and, strictly speaking, 
only two different phases. The stringy fluid phase
should have two different regimes with different properties and degrees of freedom: the stringy fluid itself above $T_{ch}$ and the
QGP at very high temperatures.

The microscopic structure of the stringy fluid
could be shortly outlined as follows:

$\bullet$ It is a densely packed system of the overlapping color-singlet
large meson-like systems.
 
$\bullet$ It is a highly collective medium with a very small mean-free
path of the color-singlet constituents.

$\bullet$ There are no deconfined gluons.

$\bullet$ The propagating in time degrees of freedom are only color-singlets,
in which the massless quarks and antiquarks are connected by the chromoelectric string into the color-singlet objects.

$\bullet$  Quark interchanges between the overlapping color-singlet meson-like systems,
required by Pauli principle, make fluctuations of conserved charges looking
as if the quarks were free \cite{Glozman:2026glk}.

\section{Conclusions}\label{sec:conclusions}

Here we summarize main points of the article.

\begin{itemize}%
\item{We have discussed a new symmetry in QCD with light quarks, the chiral spin symmetry
and its flavor extensions, which is a
symmetry of the color charge and confining electric part of QCD.}

\item{This symmetry has been observed in hadron spectrum
in the vacuum upon artificial elimination of the spontaneous
breaking of chiral symmetry on the lattice.}

\item {There is a number of implications of this observation.
It would be incorrect to say
 that the hadron mass comes from the quark condensate, i.e., is a consequence of the spontaneous breaking of chiral symmetry in the vacuum. 
 The effects of magnetic interactions of quarks with gluonic
field are at least predominantly located in the near-zero modes of the Dirac operator while the confining electric interaction is distributed in all
modes of the Dirac operator. Confinement and spontaneous breaking of chiral symmetry
are not directly related phenomena.  Still it is possible
that  confinement induces the spontaneous breaking of chiral
symmetry. However, the elimination of the chiral symmetry breaking,
e.g. in the medium at high temperatures, does not require deconfinement.}

\item { The chiral spin symmetry has been observed on the lattice at
temperatures above the chiral symmetry restoration crossover at $T_{ch}$. This
means that QCD above the chiral crossover is still in the confining regime
where propagating in time degrees of freedom are the color-singlet objects. This regime is called a
stringy fluid. The symmetry disappears at essentially higher temperature,
$T_d$, because the electric confining interaction gets screened. Consequently
above $T_d$ we have a quark-gluon plasma regime with (quasi)partons being the
effective degrees of freedom. The transition from the stringy fluid to QGP is very broad, essentially broader than the chiral crossover.}

\item {The three regimes of QCD are characterized by different scaling
of the thermodynamic quantities with $N_c$: $N_c^0$ in the hadron gas,
$N_c^1$ in the stringy fluid and $N_c^2$ in the quark-gluon plasma. These
different scalings can be observed on the lattice upon simulations
of the equation of state at $N_c=5,7$ and comparison of the results
with already known at  $N_c=3$. The physical origin of the
$N_c^1$ scaling in the stringy fluid are the fluctuations of the overlapping color singlet quark-antiquark systems. In the large $N_c$ limit the smooth
crossovers may become first order phase transitions and the three-regimes
picture of the phase diagram transforms into the three-phases at small
chemical potentials.}

\item {We have demonstrated that fluctuations of conserved charge
scale above $T_{ch}$ as $N_c^1$. Consequently a very well visible
transition on the lattice from the hadron gas at $T_{ch}$ demonstrates the
$N_c^0 \rightarrow N_c^1$ transition to the stringy fluid.}

\item {We have discussed a manifestly confining and chirally symmetric
large $N_c$ model in 3+1 dimensions that is similar to the 't Hooft
model in 1+1 dimensions. This model demonstrates the chiral symmetry
restoration in the confining regime. The reason is Paili blocking
of the quark levels necessary for the existence of the quark condensate,
by the thermal excitations of quarks. Above the chiral restoration phase transition the quark-antiquark color-singlet bound systems become
very large. Consequently the stringy fluid matter represents a highly
collective matter
of huge overlapping color-singlet quark-antiquark systems with a very
small mean free path of the color-singlet constituents. This result
very naturally explains main observations made for the hot matter 
at RHIC and LHC.}
\end{itemize}

\begin{ack}[Acknowledgments]%
The author thanks the Austrian Science Fund (FWF) for support
through the grant PAT3259224.
\end{ack}

%\bibliographystyle{elsarticle-num}
%\bibliographystyle{Harvard}
%\bibliography{reference}

\begin{thebibliography}{10}

%\cite{Banks:1979yr}
\bibitem{BC} 
Banks T. and Casher A. (1980),
%``Chiral Symmetry Breaking in Confining Theories,''
{\it Nucl. Phys.} B169:103.

%\cite{Cohen:1996sb}
\bibitem{CJ}
Cohen T. D. and Ji X. D. (1997),
{\it Phys. Rev.} D92:6870

%\cite{Glozman:2022zpy}
\bibitem{G1}
Glozman L. Ya. (2023),
{\it Prog. Part. Nucl. Phys.} 131:104049

 %\cite{Glozman:2015qva}
\bibitem{GP}
Glozman L. Ya. and Pak M. (2015),
{\it Phys. Rev.} D92:016001.

%\cite{Denissenya:2014poa}
\bibitem{D1}
Denissenya M., Glozman L. Ya. and Lang C. B. (2014),
{\it Phys. Rev.} D89:077502.

%\cite{Glozman:2014mka}
\bibitem{G2} 
Glozman L. Ya. (2015),
{\it Eur. Phys. J.} A51:27.

%\cite{Jackson:2002rj}
\bibitem{Jackson}
Jackson J. D. (2002),
{\it Am. J. Phys.} 70:917.
 
%\cite{Denissenya:2015mqa}
\bibitem{DGP} 
 Denissenya M., Glozman L. Ya. and Pak M. (2015),
{\it Phys. Rev.} D91:114512.

%\cite{Denissenya:2015woa}
\bibitem{DGP2} 
 Denissenya M., Glozman L. Ya. and Pak M. (2015),
{\it Phys. Rev.} D92:074508 [E:D92:099902].



%\cite{Bali:2000gf}
\bibitem{Bali}
 Bali G. S. (2001),
%``QCD forces and heavy quark bound states,''
{\it Phys. Rept.} 343:1.

%\cite{Denissenya:2014ywa}
\bibitem{D2}
Denissenya M, Glozman L. Ya. and Lang C. B. (2015),
{\it Phys. Rev.} D91:034505.

%\cite{Cohen:2015ekf}
\bibitem{Cohen} 
Cohen T. D. (2016),
 {\it Phys. Rev.} D93:034508.

%\cite{Christ:1980ku}
\bibitem{Lee}
Christ N. H. and Lee T. D.(1980) ,
 {\it Phys. Rev.} D22:939.

%\cite{Hagedorn:1965st}
\bibitem{Hag}
Hagedorn R. (1965),
{\it Nuovo Cim. Suppl.} 3:147.

%\cite{Cabibbo:1975ig}
\bibitem{Cab}
Cabibbo N. and Parisi G. (1975),
{\it Phys. Lett.} B59:67.
 
%\cite{Collins:1974ky}
\bibitem{Col}
Collins J. C. and Perry M. J. (1975),
{\it Phys. Rev. Lett.} 34:1353.

%\cite{Shuryak:1980tp}
\bibitem{Shuryak}
Shuryak E. V. (1980),
{\it Phys. Rept.} 61:71.

%\cite{Heinz:2013th}
\bibitem{Heinz}
Heinz U. and Snellings R. (2013),
{\it Ann. Rev. Nucl. Part. Sci.} 63:123.

%\cite{Aoki:2006we}
\bibitem{Aoki1}
Aoki Y. et al (2006),
{\it Nature} 443:675.

%\cite{Aoki:2009sc}
\bibitem{Aoki2}
Aoki Y. et al (2009),
{\it JHEP} 06:088.

%\cite{Polyakov:1978vu}
\bibitem{Polyakov}
Polyakov A. M. (1978),
{\it Phys. Lett.} B72:477.
 
 %\cite{Glozman:2016swy}
 \bibitem{G3} 
 Glozman L. Ya. (2017),
{\it Acta Phys. Polon. Supp.} 10:583.

%\cite{Rohrhofer:2019qwq}
\bibitem{R1}
Rohrhofer C. et al (2019), 
{\it Phys. Rev.} D100:014502.
 
 
%\cite{Rohrhofer:2019qal}
\bibitem{R2} 
Rohrhofer C., Aoki Y., Glozman L. Ya., and Hashimoto S. (2020),
{\it Phys. Lett.} B802:135245. 
 
%\cite{Rohrhofer:2017grg}
\bibitem{R3}
Rohrhofer C. et al (2017), 
{\it Phys. Rev.} D96:094501 [E: D99:039901].
 
%\cite{Chiu:2023hnm}
\bibitem{Chiu}
Chiu T. W. (2023),
{\it Phys. Rev.} D107;114501.

%\cite{Cohen:2023hbq}
\bibitem{CG1}
Cohen T. D. and Glozman L. Ya. (2024),
{\it Eur. Phys. J.} A60:171.

%\cite{Cohen:2024ffx}
\bibitem{CG2}
Cohen T. D. and Glozman L. Ya. (2024),
{\it Eur. Phys. J.} A60:170.

%\cite{Mickley:2024vkm}
\bibitem{Allton}
Mickley J. A., Allton C., Bignell R., and Leinweber D. B. (2025),
{\it Phys. Rev.} D111:034508.

%\cite{Lucini:2003zr}
\bibitem{Lucini}
Lucini B., Teper M. and Wenger U. (2004),
{\it JHEP} 01: 061.

%\cite{Borsanyi:2013bia}
\bibitem{Bor}
Borsanyi S. et al (2014), 
 {\it Phys. Lett.} B730:99.

%\cite{HotQCD:2014kol}
\bibitem{hot}
Bazavov A \textit{et al.} [HotQCD] (2014),
{ \it Phys. Rev.} D90:094503.

%\cite{Bazavov:2017dsy}
\bibitem{bpw}
Bazavov A., Petreczky P. and Weber J. H. (2018),
%``Equation of State in 2+1 Flavor QCD at High Temperatures,''
{\it Phys. Rev.} D97:014510.

%\cite{Glozman:2025rhe}
\bibitem{G4}
Glozman L. Ya. (2025),{\it Eur. Phys. J. } C85:1358. 

    
%\cite{tHooft:1973alw}
\bibitem{Hooft} 
't Hooft G (1974), 
{\it Nucl. Phys.} B72:461.

%\cite{Witten:1979kh}
\bibitem{Witten}
Witten E. (1979), 
{\it Nucl. Phys.} B160:57.

%\cite{Asakawa:2015ybt}
\bibitem{Asakawa}
Asakawa M. and Kitazawa M. (2016),
{\it Prog. Part. Nucl. Phys.} 90:299.

%\cite{Bellwied:2015lba}
\bibitem{Bel}
Bellwied R. et al (2015), 
{\it Phys. Rev.} D92:114505.

%\cite{Fujimoto:2025sxx}
\bibitem{ffhl}
Fujimoto Y., Fukushima K., Hidaka Y. and McLerran L. (2025),
{\it Phys. Rev.} D112:074006.

%\cite{LeYaouanc:1983huv}
\bibitem{LeYaouanc:1983huv}
Yaouanc A. Le. et al (1984), 
{\it Phys. Rev.} 
  D29:1233.

%\cite{LeYaouanc:1984ntu}
\bibitem{LeYaouanc:1984ntu}
Yaouanc A. Le. et al (1985), 
{\it Phys. Rev.} 
  D31:137.

%\cite{Adler:1984ri}
\bibitem{Adler:1984ri}
Adler S. L. and  Davis A. C. (1984), 
{\it Nucl.
  Phys.}  B244:469.

%\cite{Kocic:1985uq}
\bibitem{Kocic:1985uq}
Kocic A. (1986), 
{\it Phys.
  Rev. } D33:1785.

%\cite{Bicudo:1989sh}
\bibitem{Bicudo:1989sh}
 Bicudo P. and  Ribeiro J. (1990), 
 {\it Phys.
  Rev. } D42:1611.

%\cite{Bicudo:1989si}
\bibitem{Bicudo:1989si}
Bicudo P. and Ribeiro J. (1990), 
{\it Phys. Rev.} D42:1625.

%\cite{Llanes-Estrada:1999nat}
\bibitem{Llanes-Estrada:1999nat}
 Llanes-Estrada F. J. and Cotanch S. R. (2000), 
 {\it Phys. Rev. Lett.} 84:1102.
  
%\cite{Wagenbrunn:2007ie}
\bibitem{Wagenbrunn:2007ie}
Wagenbrunn R. F. and Glozman L. Ya. (2007), 
{\it Phys. Rev.} D75:036007. 

%\cite{Quandt:2018bbu}
\bibitem{reinhardt}
Quandt M., Ebadati E., Reinhardt H. and Vastag P. (2018), 
{\it Phys. Rev.} D98:034012.


%\cite{Glozman:2024xll}
\bibitem{gnw1}
 Glozman L. Ya.,  Nefediev A. V. and Wagenbrunn R. (2024), 
 {\it Phys. Lett. }
  B854:138707.



%\cite{Glozman:2024dzz}
\bibitem{gnw2}
 Glozman L. Ya.,  Nefediev A. V. and Wagenbrunn R. (2025), 
 {\it Eur. Phys. J. } C85:462.

%\cite{tHooft:1974pnl}
\bibitem{tHooft:1974pnl}
't Hooft G. (1974), {\it Nucl. Phys.}
  B75:461.


%\cite{Bars:1977ud}
\bibitem{Bars:1977ud}
Bars I. and  Green M. B. (1978), 
{\it Phys. Rev. } 17:537.

%\cite{Kalashnikova:2001df}
\bibitem{Kalashnikova:2001df}
Kalashnikova Y. S. and  Nefediev A. V. (2002), 
{\it Phys. Usp.} 45:347. 

%\cite{Glozman:2012ev}
\bibitem{Glozman:2012ev}
 Glozman L. Ya.,  Sazonov V. K., Shifman M. and Wagenbrunn R. F. (2012), {\it Phys. Rev.} 
  D85:094030.

%\cite{HotQCD:2019xnw}
\bibitem{kkk}
 Ding H. T. et~al. (2019), {\it Phys. Rev.
  Lett.}  123:062002.
 
 %\cite{YaGlozman:2026ywk}
\bibitem{YaGlozman:2026ywk}
Glozman L. Ya. (2026), {\it Phys. Lett.} B883:141099.

%\cite{Glozman:2026glk}
\bibitem{Glozman:2026glk}
Glozman L. Ya. (2026), {\it Eur. Phys. J.} C86:1223.

  
\end{thebibliography}

\end{document}